\documentclass[english]{article}
\usepackage[T1]{fontenc}
\usepackage[latin9]{inputenc}
\usepackage{amsmath}
\usepackage{amssymb}
\usepackage{fullpage}
\usepackage{float}
\usepackage{natbib}
\usepackage{graphicx}
\usepackage{tabularx}
\usepackage{multirow}
\usepackage{hyperref}
\makeatletter

\providecommand{\tabularnewline}{\\}

\usepackage{fullpage}

\makeatother

\usepackage{babel}

\begin{document}

\setcitestyle{numbers,square}

\title{Effective Genetic Risk Prediction Using Mixed Models}

\author{David Golan and Saharon Rosset}

\maketitle

\section*{Abstract}
To date, efforts to produce high-quality polygenic risk scores from genome-wide studies of common disease have focused on estimating and aggregating the effects of multiple SNPs.
Here we propose a novel statistical approach for genetic risk prediction, based on random and mixed effects models. Our approach (termed GeRSI) circumvents the need to estimate the effect sizes of numerous SNPs by treating these effects as random, producing predictions which are consistently superior to current state of the art, as we demonstrate in extensive simulation. 
When applying GeRSI to seven phenotypes from the WTCCC study, we confirm that the use of random effects is most beneficial for diseases that are known to be highly polygenic: hypertension (HT) and bipolar disorder (BD). 
For HT, there are no significant associations in the WTCCC data. The best existing model yields an AUC of 54\%, while GeRSI improves it to 59\%. For BD, using GeRSI improves the AUC from 55\% to 62\%. 
For individuals ranked at the top 10\% of BD risk predictions, using GeRSI substantially increases the BD relative risk from $1.4$ to $2.5$. 

\section*{Introduction}

Despite the huge investment and considerable progress in the study
of the genetic causes of human diseases, the underlying genetic mechanisms of many common diseases, including type-1 diabetes, bipolar disorder, schizophrenia, multiple sclerosis and Alzheimer\textquoteright{}s disease, are still largely unknown. These diseases are known to be highly heritable, based on family studies, and so it was expected that the \textquotedblleft{}genetic revolution\textquotedblright{} would considerably advance the understanding and treatment of these diseases.

The leading methodology for finding the genetic causes of disease
is the Genome-Wide Association Study (GWAS). In a typical GWAS, one
collects thousands of sick and healthy individuals, genotypes them,
and searches for single-nucleotide polymorphisms (SNPs) which are
more abundant in one group or the other. To date, GWASs have flagged thousands of SNPs as associated with hundreds of diseases. However, our ability to accurately predict an individual\textquoteright{}s disease status based on these SNPs still falls considerably short of what\textquoteright{}s expected based on the high heritability of these diseases. 

This remarkable gap between the predictive power of significantly associated SNPs and the expected predictive capacity based on the
high heritability of the phenotypes has been termed the {}``mystery'' or {}``problem'' of the missing heritability. One leading theory
attempting to explain this mystery is that many phenotypes are driven by a plethora of common SNPs with small effects, and that current day GWASs are underpowered to detect these SNPs due to their small effects. Goldstein (2009) \cite{goldstein2009common} estimated the overall number of SNPs affecting height at 93,000. 

In light of this theory, the traditional naive approach of using only the SNPs that were found to be significantly associated with the disease in calculating genetic risk scores (GRSs) is expected to perform poorly, as it overlooks the lion's share of causal SNPs, whose effect is not large enough to be declared significant. Instead, recent efforts in computing GRSs have attempted to include a larger number of SNPs, primarily by adopting much more lenient inclusion criteria for SNPs. 

Intuitively, using a more permissive threshold has two results: the inclusion of more {}``true'' causal SNPs, which improve the quality of the genetic risk prediction, and the inclusion of more {}``false'', or non-causal SNPs, which add noise to the genetic risk prediction. A good choice of p-value threshold would be such that the trade-off between signal and noise is beneficial. Recent studies demonstrated that GRSs computed in this manner have a significant association with the disease, above and beyond GRSs computed using only significantly associated SNPs, when including up to half of the genotyped SNPs (see, e.g., \cite{purcell2009common}). These results suggest that for at least some diseases there's considerable information in the long tail of insignificantly associated SNPs, and that the benefit from including more true positives trumps the cost of including more false positives. Dudbridge (2013) \cite{dudbridge2013power} and Chatterjee et al. (2013) \cite{chatterjee2013projecting} provide an in depth mathematical analysis of this approach and variants thereof, and it was recently applied for prediction of risk of Celiac disease with remarkable success \cite{abraham2013accurate}. 

Most of the methods for computing GRSs fall under the same category which is known in the statistics literature as {}``fixed effects'' modeling. In such models the effects of SNPs are assumed to be parameters (i.e. fixed, but unknown, quantities). These parameters are estimated and used in subsequent analysis. For example, one would estimate the odds-ratio (OR) of a given SNP from a GWAS, and use this estimate to predict the risk of new individuals. The main difference between the methods lies in the way these parameters are estimated, ranging from simple SNP-by-SNP regression, to shrinkage based estimates such as the LASSO and more, as reviewed by \cite{dudbridge2013power,abraham2013accurate}. They also differ in the way SNPs are chosen to be included in computation of the risk scores \cite{dudbridge2013power,abraham2013accurate}. However, they all share the fundamental treatment of the effects as parameters which require estimation. 

In this manuscript we adopt an alternative approach to computing GRSs, known as ``random effects'' modeling. The basic premise of this approach is that our goal is not to estimate the individual effect of every SNP, but rather their accumulated effect. With a large number of causative SNPs, the number of parameters to estimate in a fixed effects approach can become very large, resulting in poor statistical estimation properties, which in turn reduces the ability to correctly predict the phenotype of out-of-sample individuals. 

Hence, it might be beneficial to circumvent estimating each and every effect. To accomplish this goal, effect sizes are treated not as parameters, but rather as random variables with some prior distribution. They can then be {}``integrated out'', thus mitigating the need to estimate them separately. Instead, a correlation (or kinship) matrix $G$ is estimated using the genotypes, and models the correlations between the GRSs. The correlation between GRSs of individuals who are more similar genetically would be higher, and vice-versa. 

The random effects approach has been adopted in the context of GWAS for association tests \cite{lippert2011fast,kang2010variance,price2010new,zhou2012genome,zhou2014efficient} and heritability estimation \cite{yang2010common,lee2011estimating,zhou2013polygenic,golan2011accurate} with much success. All of these approaches rely on treating the phenotype as a normally distributed variable, which is a sum of a genetic and an environmental component,  
and utilizing well established linear mixed models (LMM) methodologies to draw inferences about their quantities of interest. 

In addition, the animal-breeding literature uses similar approaches to model and estimate the ``breeding-value'', which is closely related to the genetic risk. In the scenario of an observational study of a quantitative phenotype, there is a well established methodology for estimation of the breeding-value, known as the best linear unbiased predictor, or BLUP \cite{vanraden2008efficient,hayes2001prediction}. When estimating breeding values, pedigree data are usually available. However, the kinship can be estimated from genotype data (referred to as genetic-BLUP or gBLUP \cite{clark2013genomic}) and this method is implemented in the widely used GCTA software \cite{yang2011gcta,clark2013genomic}. 

However, case-control studies present a much more challenging statistical setup. First, the phenotype is binary rather than quantitative, and so cannot be accurately modeled by a multivariate normal distribution. 
Additionally, cases are highly over-represented in the sample, compared to the population, and so many of the typical statistical assumptions (namely normality  and independence of the genetic and environmental effects) are no longer legitimate. 

A common approach to random effect modeling in case-control GWAS is to treat the phenotype as quantitative, and apply LMM methodologies, possibly followed by post-hoc corrections to account for violation of the underlying assumptions \cite{zhou2013polygenic,lee2011estimating}. While this approach has proven successful in practice, its reliance on probabilistic models which are known to be inaccurate is expected to result in sub-optimal performance. In the context of GRS estimation, the natural extension of the LMM approach is to use gBLUP, but this is subject to similar inaccuracy concerns, as our simulations below demonstrate. 

We describe a novel approach for Genetic Risk Scores Inference (GeRSI). GeRSI is based on a Markov-chain Monte-Carlo (MCMC) method utilizing Gibbs sampling to estimate the GRSs of individuals given the genotypes of a case-control study. We use the well known liability threshold model to model the dichotomous nature of the phenotype. Additionally, our Gibbs sampling approach conditions explicitly on the selection of individuals to the study, and thus accounts directly for the over-representation of cases in the study. By properly conditioning on the selection, we can sample from the true posterior distribution of the GRS. This is in contrast to using LMM-based approaches, which treat a case-control disease phenotype as if it were a randomly sampled quantitative one.

In addition to accounting for disease phenotypes and non-random selection in prediction, GeRSI can be naturally extended to accommodate fixed effects within the probabilistic framework. Hence our approach also allows ``mixed effects'' modeling, where SNPs with considerable effects can be included as fixed effects, while the long tail of insignificantly associated SNPs is accounted for using random effects. We distinguish between random-effect GeRSI, which treats all SNPs as random effects, and mixed-effects GeRSI, which includes SNPs below a certain p-value threshold as fixed effects, treating the rest of the SNPs are random-effects. 
Additionally, introducing fixed effects to the model allows accounting for additional covariates such as sex, ethnicity, and known environmental risk factors (e.g. smoking habits). 


\section*{Results}
We tested GeRSI in extensive simulations (see Online Methods for details). Prediction quality is measured by the area under the ROC curve (AUC), which is the probability that a randomly sampled affected individual (case) attains a higher genetic risk score than a randomly sampled unaffected individual (control). 

Our results demonstrate, as expected, that fixed effects modeling is generally effective when the phenotype is driven by a small number of SNPs with sizable effects (i.e., most causal SNPs are easy to identify), and that the random effects approach is most effective when the phenotype is driven by a large number of SNPs with small effects. The hybrid mixed-effects GeRSI performs well in both scenarios, as well as in intermediate scenarios, and is never inferior to fixed effects modeling or gBLUP in all our simulations. 

\begin{figure}[H]
\centering
\includegraphics[scale=0.6]{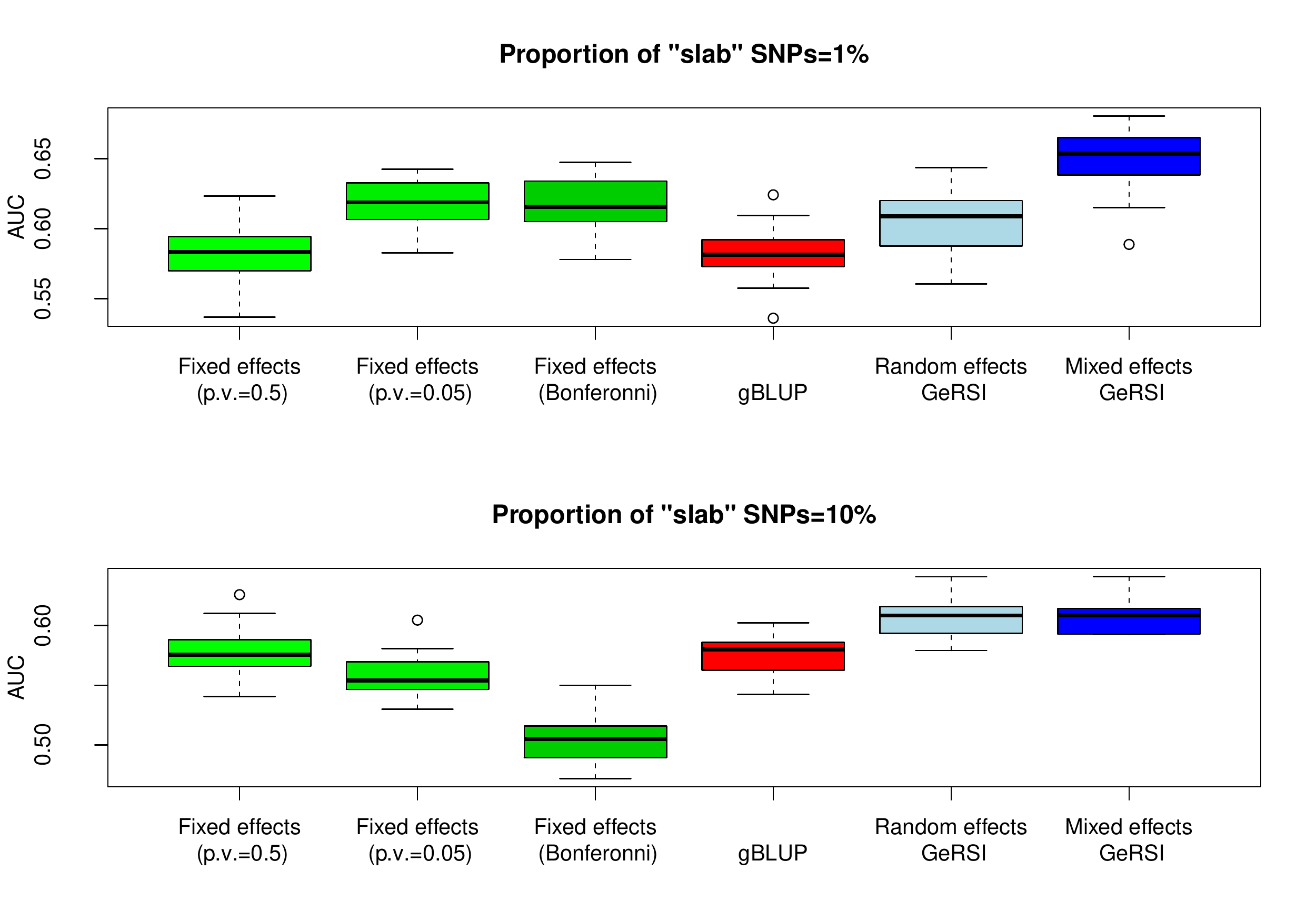}
\caption{\textbf{Comparison of the performance of fixed, random and mixed effects models in disease risk prediction in a spike and slab model.} We simulated balanced case-control studies of a disease with prevalence 5\% and heritability 50\%, where the fraction of "slab" SNPs with large effects was either $1\%$ or $10\%$ out of a total of 50,000 simulated SNPs, and these slab SNPs account for $90\%$ of the heritability in line with values from \cite{chatterjee2013projecting}. We show the performance of the fixed-effects approach with (a) Bonferroni adjusted p-value threshold (b) p-value threshold of $0.05$, and (c) p-value threshold of $0.5$. In addition we compute the correlation matrix $G$ and use it to predict risk using random-effects GeRSI approach, as well as mixed-effects GeRSI treating the SNPs from (a) as fixed-effects. We also show the performance of the standard gBLUP random effects approach. In each simulation, we used a train set of $3,000$ individuals, and estimated the AUC for each method using a test set of $1,000$ individuals. Results from 20 independent simulations were used to draw the boxplots.}
\label{mix_fig}

\end{figure}

In Figure \ref{mix_fig} we present the results for the "spike and slab" model of genetic risk, recently explored by Zhou et al. (2013) and Chatterjee et al. (2013) \cite{zhou2013polygenic,chatterjee2013projecting}, and found to provide a good fit for the observed effect sizes for a wide variety of GWAS. In this model, all SNPs have effects on the phenotype, but this is divided between a small fraction of ``slab'' SNPs with considerable effect sizes, and the bulk of ''spike'' SNPs with very small, but non-zero effects. The results illustrate the power and flexibility of GeRSI, and its superior performance compared to both fixed effects modeling and gBLUP (superior in all $20$ simulation runs, for both setups, p-value$<10^{-6}$, using sign-tests). When slab effects are relatively small (as in the bottom panel), random effects GeRSI and mixed effects GeRSI perform similarly, but in the presence of large slab effects (as in the top panel), the mixed effects version allows us to capture these as fixed effects and is far superior to the random effects version. 
The results for other simulation settings are presented in Supplementary Figures 1-21.
Importantly, random-effects GeRSI trumps gBLUP in all our simulations, as expected due to the fact that it utilizes the correct probabilistic model rather than an approximated model. 

We then proceeded to apply GeRSI to seven case-control studies from the Wellcome Trust Case-Control Consortium (WTCCC)\cite{WTCCC}: Bipolar disorder (BD), Coronary artery disease (CAD), Crohn's disease (CD), hyper-tension (HT), Type-1 and Type-2 diabetes (T1D and T2D, respectively) and Rheumatoid arthritis (RA). For each phenotype, we first performed stringent quality control as suggested by Lee et al. \cite{lee2011estimating} to mitigate batch effects. We then estimated the AUC with four-fold cross validation using both the fixed effects method of \cite{dudbridge2013power} and the random-effects and mixed-effects GeRSI approach. To demonstrate the potential clinical utility of the risk prediction approaches examined, we also estimated the relative risk (RR) of an individual found to be in the top 1\% and 10\% of risk predictions. For the fixed-effects approaches, we considered a range of possible p-value thresholds:  $5\times 10^{-c}$ for $c \in \{1,...,8\}$, and display here the best result for each phenotype. 

\begin{figure}
\begin{minipage}[t]{0.45\columnwidth}%
\includegraphics[scale=0.4]{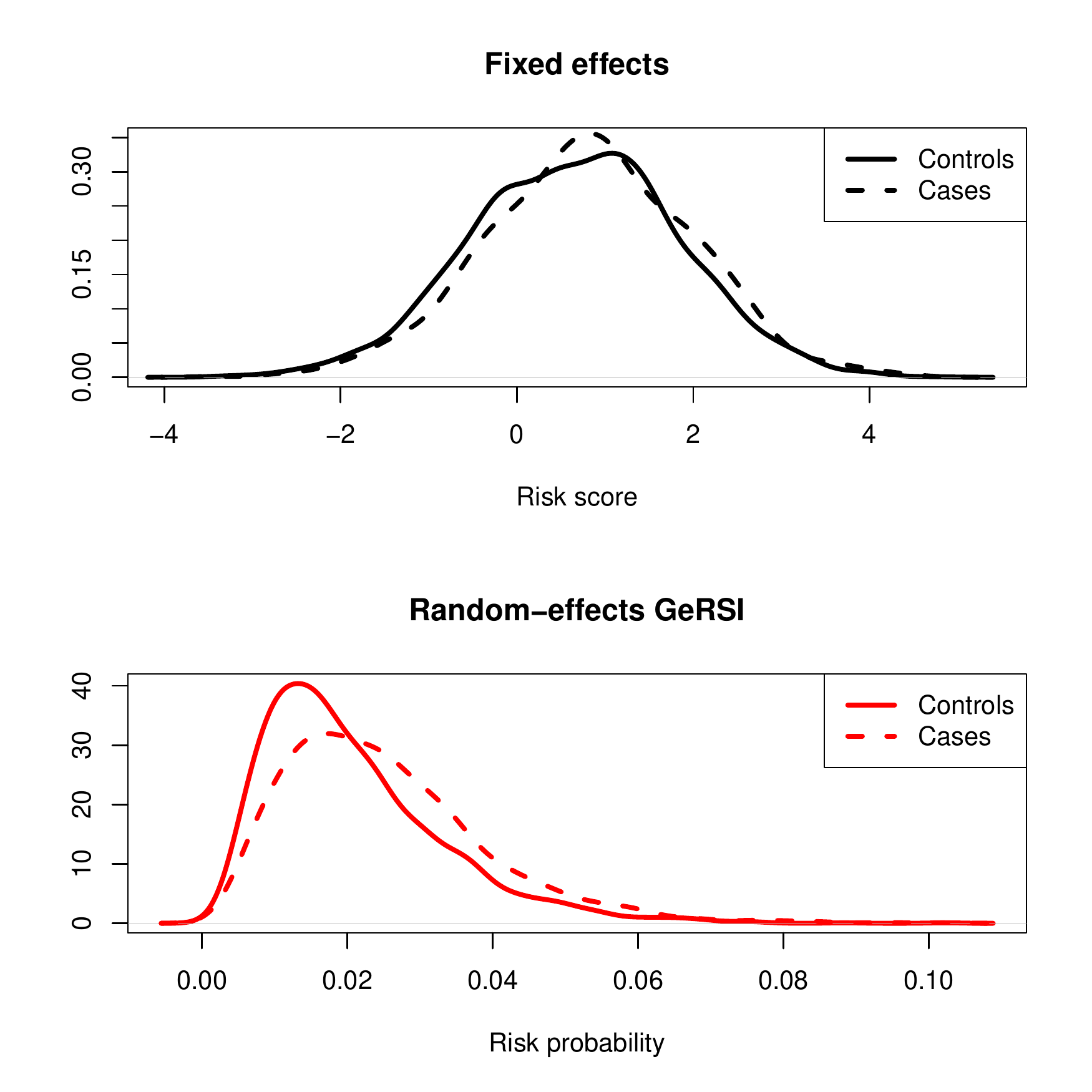}%
\end{minipage}\hfill{}%
\begin{minipage}[t]{0.45\columnwidth}%
\includegraphics[scale=0.4]{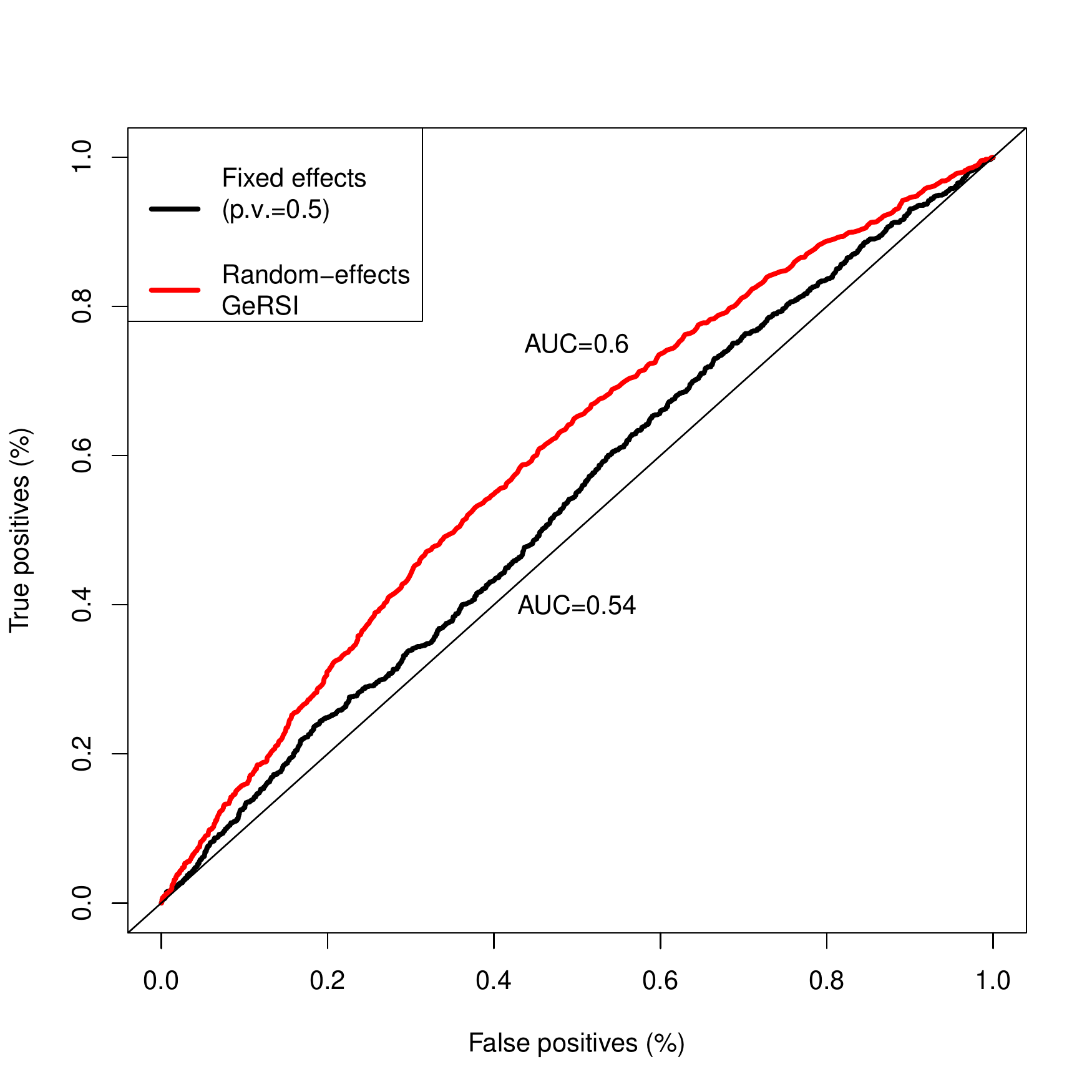}%
\end{minipage}

\caption{{\bf Comparison of hypertension risk predictions using fixed effect models and random-effects GeRSI}. We used the fixed-effects approach with a p-value threshold of $0.5$. When using fixed effects, there's very little difference between the distribution of risk scores of cases and control (top-left panel), but when using random effects GeRSI, out-of-sample risk predictions for cases are clearly skewed to the right (bottom-left panel). This is also evident when comparing the ROC curves of both methods (right panel).}

\label{HT_compare}
\end{figure}

Comparing the fixed effects approach to random-effects GeRSI, we observed that random-effects GeRSI obtained significantly higher AUC than the optimized fixed effects approach for four of the seven phenotypes: BD, T2D, CAD and HT (Table \ref{tab1}, p-value$<10^{-9},10^{-3},10^{-3},10^{-5}$, respectively. See Supplementary Materials for details). Specifically for HT, there are no SNPs found to be associated at the $5\times 10^{-8}$ genome-wide significance level in the WTCCC data \cite{WTCCC}, and very few associations have been found in any studies to date (only seven associations are listed in the NHGRI GWAS catalog \cite{NHGRI}). The optimal p-value threshold for fixed-effects model was $0.5$, indicating a true polygenic architecture, and thus an ideal phenotype for random-effects modeling. As expected, random-effects GeRSI yields substantially better predictive power, with AUC around $0.6$ and a considerable increase in relative risk for individuals at the top 10\% of GeRSI risk predictions ($1.61$ compared to $1.31$). Figure \ref{HT_compare} contrasts the ROC curves and the risk prediction behavior of the two approaches on HT data. 

Contrary to HT, there are numerous replicated associations with BD, some of which were identified in the original WTCCC study. However, in agreement with other studies \cite{purcell2009common}, we find that using a permissive p-value threshold for risk predictions is appropriate (the optimal p-value threshold was $0.5$). 
Here, too, using random-effects modeling is beneficial, as expected. This is reflected in the AUC and 10\% relative risk numbers in Table \ref{tab1}. Additionally, the top 1\% of risk scores attain a relative risk of almost $4$ in random-effects GeRSI, compared to $1.26$ using the fixed-effects approach (see Supplementary Materials). 

Random-effects GeRSI did not improve over fixed effects modeling for CD, and performed significantly worse for RA and T1D (Table \ref{tab2}). This is consistent with our knowledge regarding the genetic architecture of these diseases: all three are auto-immune diseases, with strongly associated SNPs with considerable effect sizes, primarily in the MHC region on chromosome 6. Gusev et al. (2014) \cite{gusev2013quantifying} recently showed that a significant portion of the heritability of these diseases is due to variants in the vicinity of previously identified causal SNPs, and is not uniformly distributed along the genome. 
To demonstrate the flexibility of mixed-effects GeRSI to combine the power of fixed and random effects modeling, we also show its performance in Table \ref{tab2}, where it is generally comparable to that of the fixed effects approach (slightly superior for CD, p-value$<10^{-3}$).

\begin{table}
\centering
\begin{tabular}{|c||c|c|c||c|c|}
\hline
 \multirow{2}{*}{Phenotype} &
 \multicolumn{3}{c||}{ Fixed} &
 \multicolumn{2}{c|}{ Random} \\
\cline{2-6}
 & Best thresh. & AUC (s.e.) & RR at 10\% [CI] & AUC (s.e.) & RR at 10\% [CI]  \\
\hline 
\multirow{2}{*}{BD}  &  \multirow{2}{*}{0.5} & 0.55 & 1.4 & 0.62 & 2.5  \tabularnewline 
 
 & & (0.01) & [1.03-1.65] & (0.01) & [2.16-2.96] \tabularnewline 
 \hline

\multirow{2}{*}{T2D}  &  \multirow{2}{*}{0.005} & 0.55 & 1.48 & 0.59 & 1.67  \tabularnewline 
 
 & & (0.01)  & [1.2-1.83] & (0.01) & [1.34-1.99] \tabularnewline 
 \hline

\multirow{2}{*}{CAD}  &  \multirow{2}{*}{5e-05} & 0.65 & 1.85 &  0.67  & 2.16  \tabularnewline 
 
 & & (0.01) & [1.17-2.3] &  (0.01) & [1.79-2.72] \tabularnewline 
 \hline

\multirow{2}{*}{HT}  &  \multirow{2}{*}{0.5} & 0.54 & 1.31  & 0.60 & 1.61  \tabularnewline 
 
 & & (0.01) & [1.06-1.55] & (0.01) & [1.42-1.97] \tabularnewline 
 \hline 

\hline \end{tabular}
\caption{{\bf Comparison of the Fixed-effects approach and random-effects GeRSI on four phenotypes from the WTCCC}. Estimated AUC and RR of the top 10\% and top $1\%$ of individuals, estimated using four-fold cross-validation. We compare the predictions obtained from our random-effects GeRSI approach to the predictions obtained using a fixed-effects approach.
We compute the AUC of the fixed-effects approach for a wide range of p-value thresholds: $5\times10^{-c}$ for $c\in{1,...,8}$, and display here results for the value with the highest AUC (note that GeRSI has no such parameter).
All analyses include sex as a covariate.}
\label{tab1}
\end{table}

\begin{table}
\begin{center}
\begin{tabular}{|c||c|c|c||c|c|c|}
\hline
 \multirow{2}{*}{Phenotype} & 
\multicolumn{3}{c||}{ AUC (s.e.)} &
\multicolumn{3}{c|}{ RR at top 10\% [CI]} \\
\cline{2-7}
  & Fixed & Random & Mixed & Fixed & Random & Mixed \\
\hline 
\multirow{2}{*}{CD}  & 0.59 & 0.59 & 0.62 & 2.02 & 2.18 & 2.63 \tabularnewline 
 
 & (0.01) & (0.01) & (0.01) & [1.67-2.34] & [2.12-2.53] & [2.32-2.81]\tabularnewline 
 \hline 
\multirow{2}{*}{T1D}  & 0.72 & 0.55 & 0.71 & 3.4 & 1.46 & 3.45 \tabularnewline 
 
 & (0.01) & (0.01) & (0.01) & [3.19-3.63] & [1.31-1.83] & [2.76-3.57]\tabularnewline 
 \hline 
\multirow{2}{*}{RA}  & 0.67 & 0.63 & 0.68 & 2.85 & 1.83 & 2.93 \tabularnewline 
 
 & (0.01) & (0.01) & (0.01) & [2.62-3.61] & [1.44-2.07] & [2.55-3.54]\tabularnewline 
 \hline 
 
 \hline \end{tabular}
\end{center}
\caption{{\bf Comparison of the random and mixed effects GeRSI approach and the fixed effects approach for three autoimmune diseases in the WTCCC data.} Estimated AUC and RR of the top 10\% of individuals, estimated using four-fold cross-validation. We compare the predictions obtained from our random-effects GeRSI approach, our mixed-effects GeRSI approach and predictions obtained using a fixed-effects approach. 
We compute the AUC of the fixed-effects approach and the mixed-effects GeRSI approach for a wide range of p-value thresholds: $5\times10^{-c}$ for $c\in{1,...,8}$, and display here results for the value with the highest AUC.
All analyses include sex as a covariate.}

\label{tab2}
\end{table}

\begin{table}
\begin{center}
\begin{tabular}{|c||c|c||c|c|}
\hline
 \multirow{2}{*}{Phenotype} & 
\multicolumn{2}{c||}{ AUC (s.e.)} &
\multicolumn{2}{c|}{ RR at top 10\% [CI]} \\
\cline{2-5}
  & 0 PCs & 10 PCs & 0 PCs & 10 PCs \\
\hline 
\multirow{2}{*}{BD} & 0.62 &  0.59 & 2.5 & 1.81  \tabularnewline 
 
& (0.01) & (0.01) & [2.3-3.05] & [1.66-2.23] \tabularnewline 
 \hline 

\multirow{2}{*}{CD} & 0.59 &  0.57 & 2.18 & 1.74  \tabularnewline 
 
& (0.01) & (0.01) & [1.83-2.55] & [1.49-2.01] \tabularnewline 
 \hline 

\multirow{2}{*}{T1D} & 0.55 &  0.52 & 1.46 & 1.03  \tabularnewline 
 
& (0.01) & (0.01) & [1.09-1.7] & [0.94-1.16] \tabularnewline 
 \hline 

\multirow{2}{*}{T2D} & 0.59 &  0.57 & 1.67 & 1.66  \tabularnewline 
 
& (0.01) & (0.01) & [1.58-1.8] & [1.46-1.76] \tabularnewline 
 \hline 

\multirow{2}{*}{CAD} & 0.67 &  0.68 & 2.16 & 2.5  \tabularnewline 
 
& (0.01) & (0.01) & [1.75-2.61] & [2.24-2.93] \tabularnewline 
 \hline 

\multirow{2}{*}{RA} & 0.63 &  0.6 & 1.83 & 1.77  \tabularnewline 
 
& (0.01) & (0.01) & [1.47-1.91] & [1.45-1.97] \tabularnewline 
 \hline 

\multirow{2}{*}{HT} & 0.59 &  0.6 & 1.61 & 1.68  \tabularnewline 
 
& (0.01) & (0.01) & [1.48-1.91] & [1.46-1.85] \tabularnewline 
 \hline 

 \hline \end{tabular}
\end{center}
\caption{{\bf Investigating the possible effect of sampling-induced population structure on risk prediction accuracy estimation.} We compare the AUCs and RRs for the top $10\%$ of individuals, when using the correlation matrix $G$ directly, or after removing the top ten principal components as described in Online Methods. As expected, we observe a minor decrement in performance in some phenotypes, but others, such as CAD and HT, display no real change in performance. 
All analyses include sex as a covariate.}

\label{tab3}
\end{table}

\section*{Discussion}

An important aspect in applying prediction methodology to case-control data is that of population structure. It is well established that population structure must be accounted for in GWAS, and this is often done using mixed-models \cite{price2010new}, or by inclusion of several top principal components as covariates \cite{price2006principal}. In the context of heritability estimation, population structure can inflate the estimated heritability if unaccounted for, and typically a number of principal components of $G$ are included as fixed effects to control for that structure.
When predicting risk, the role of population structure is more complicated. We distinguish between two types of population structure: actual and induced. Actual population structure is structure which is truly present in the population and is properly reflected in the study group. Importantly, taking advantage of this type of structure for the purpose of risk prediction is legitimate, even if the effect of the structure on the phenotype is not via genetics. For example, if the diet of individuals of a certain ethnicity affects their disease risk, but this is not accounted for using fixed effects (e.g., if dietary information was not collected), this can still be captured in GeRSI via the genetic differences between these individuals and others in the population. 

On the other hand, induced structure is an artifact of the sampling procedure. For example, a certain sub-population may be considerably more likely to be sampled in the cases compared to the controls. In this case, GeRSI predictive power estimates based on the study sample might be illegitimately inflated if this structure is not accounted for. 

The WTCCC studies are considered to have relatively little structure \cite{WTCCC}, and we are not able to separate the structure that does exist into its ``legitimate'' and ``induced'' components. To examine the robustness of our results to removal of structure, we re-ran our analyses while removing the top $10$ principal components from the correlation matrix (Table \ref{tab3}). This has a small negative effect on the performance of GeRSI for some of the phenotypes, like BD and RA, but the general spirit of the results remains unchanged. 

Another key point is that of quality-control (QC). Following Lee et al. (2011), we applied very stringent QC (Online Methods). Such stringent QC is particularly important when evaluating predictions, as different cohorts were genotyped in different times and different centers, and so systematic genotyping errors manifest as inflated estimates of the predictive capacity. On the other hand, stringent QC results in fewer SNPs and fewer individuals in the inspected datasets, which may result in conservative estimates of the predictive power. 


In conclusion, GeRSI is a method which takes advantage of the power of random effects modeling to accumulate evidence from the entire genome for the purpose of obtaining accurate risk predictions from GWAS. This is accomplished through an appropriate probabilistic inference approach, also allowing for inclusion of relevant fixed effects, including associated SNPs and other covariates. Thus, the GeRSI mixed-effects approach subsumes current fixed-effects methods and can be used in all scenarios. Our results demonstrate the significant benefits of taking this approach on both simulated and real data.
Specifically, for bipolar disorder, random effects GeRSI allows us to identify 1\% of the population at four-fold risk of disease, and 10\% of the population at 2.5-fold risk. These numbers represent a major improvement over current state of the art, and bring us closer to the ultimate goal of obtaining clinically useful risk predictions from GWAS data. We believe that random effects modeling is a key component in this quest.

\part*{Acknowledgements} 

This study makes use of data generated by the Wellcome Trust Case Control Consortium. A full list of the investigators who contributed to the generation of the data is available from www.wtccc.org.uk. Funding for the project was provided by the Wellcome Trust under award 076113.

\part*{Online Methods}

{\bf Generative model of a polygenic disease.} A polygenic quantitative trait $y$ is typically modelled using the following additive model:
\[
y_i = \mu + \sum_{j\in{\cal C}}z_{ij}u_{j} + e_i,
\]
where ${\cal C}$ is the set of causal SNPs, $u_i$ is the effect of the $i$th causal SNP, $e_i$ is the environmental effect associated with individual $i$, and $z_{ij}$ is the genotype of the $j$th SNP of the $i$th individual. The term $\sum_{j\in{\cal C}}z_{ij}u_{j}$ is often referred to as the genetic effect and denoted $g_i$. Under mild independence assumptions we have $\sigma_g^2 = Var(g) = |{\cal C}|\sigma_u^2$.

We note that the choice to use standardized SNPs rather than just centering the SNPs hides an implicit assumption that SNPs with lower frequencies have larger effect sizes (as is noted in \cite{zhou2013polygenic}). However, since we focus on common SNPs, the effects of this assumption are minimal. 

Polygenic disease phenotypes are modeled using the liability
threshold model \cite{lee2011estimating}. We assume the existence of a latent quantitative {}``liability'' phenotype. If an individual's liability exceeds a certain threshold, she is assumed to be a case. The liability is modeled as a quantitative trait with mean $0$ and variance $1$. Under these assumptions the threshold corresponding to a disease prevalence $K$ is $\Phi^{-1}(1-K) = Z_{1-K}$ where $\Phi$ is the cumulative distribution function of the standard normal distribution and $\Phi^{-1}$ is its inverse (percentiles of the distribution), often denoted $Z$.\\

{\bf Random effects modeling.} We are interested in predicting risk, and the actual values of the
$u_{i}$'s are not of direct interest. We therefore model them as random
variables drawn from a distribution with mean $0$ and variance $\frac{\sigma_{g}^{2}}{|{\cal C}|}$. Typically one assumes a normal distribution (e.g., \cite{yang2010common,zhou2013polygenic}), but other distributions were suggested as well as mixture distributions \cite{chatterjee2013projecting,zhou2013polygenic}. We note that as long as the number of causal SNPs is large enough and the effect sizes are independent of each other, the genetic effect approximately follows a normal distribution, regardless of the underlying distribution of the effect sizes by virtue of the central-limit theorem. 

Assuming normality of the random effects and random-sampling, this implies the following distribution of the liabilities:

\[
l\sim MVN(0;G\sigma_{g}^{2}+I\sigma_{e}^{2}),\]
where $G$ is the correlation matrix of the genetic effects and is given by:
\[
G_{ij} = \frac{1}{|{\cal C}|}\sum_{k\in{\cal C}} z_{ik}z_{jk}.
\]
Since the identity of the causal SNPs is unknown, and additionally they are often not genotyped, we follow previous works \cite{yang2010common,lee2011estimating} and estimate
$G$ using all genotyped SNPs and use this estimate throughout. This matrix is often referred to as the observed kinship matrix. The estimation of $G$ is the subject of much recent research, but is out of the scope of the current paper.\\

{\bf GeRSI sampling scheme.} Assume we have a group of $n$ individuals with known genotypes, but the phenotypes are known only for the first $n-1$ individuals. We are interested in predicting the genetic risk of the $n$th individual. 

We denote $g$ and $e$ the vectors of latent genetic and environmental effects, respectively. The heritability (and hence $\sigma_{g}^{2}$) is assumed to be known, and in practice can be estimated directly from the data \cite{yang2010common,lee2011estimating} or obtained from family studies.

Our goal is to predict $P(l_{n}>t),$ conditional on our entire
data (namely the genotypes of all $n$ individuals and the phenotypes of the first $n-1$ individuals). Had we known $g_{n}$,
the (optimal) risk prediction $r_{n}$ under the model would have been:

\[
r_{n}=P(l_{n}>t\mid g_{n})=1-\Phi(\frac{t-g_{n}}{\sigma_{e}}).\]

However, since $g_{n}$ is unknown, we generate samples $k$ samples,
$g_{n,1},...,g_{n,k}$, from the posterior distribution of $g_{n}$,
conditional on all the observed data, and estimate the risk as:

\[
\hat{r}_{n}=\frac{1}{k}\sum_{i=1}^{k}\Big(1-\Phi(\frac{t-g_{n,i}}{\sigma_{e}})\Big).\]

To generate samples from the posterior distribution of $g_{n}$, we
note that:
\[
P(g_{n}\mid y_{-n}\;;\; G,\sigma_{g}^{2})=P(g_{n}\mid g_{-n},y_{-n}\;;\; G,\sigma_{g}^{2})P(g_{-n}\mid y_{-n}\;;\; G,\sigma_{g}^{2})=\]

\[
=P(g_{n}\mid g_{-n}\;;\; G,\sigma_{g}^{2})P(g_{-n}\mid y_{-n}\;;\; G,\sigma_{g}^{2}).\]
In other words, sampling the posterior can be decomposed into two
separate problems. The first problem is the problem of sampling $g_{n}$
given the values of the other genetic effects $g_{-n}.$ Since we
are conditioning on the genetic effects, $g_{n}$ is independent of
the phenotypes of the other individuals.

As we show in the Supplementary Materials, even when the sampling is not random (e.g., in case-control studies), the conditional distribution of $g_{n}$ is given by:
\[
g_{n}|g_{-n};G,\sigma_{g}^{2}\sim MVN(G_{i,-i}G_{-i,-i}^{-1}g_{-i},G_{n,-n}G_{-n,-n}^{-1}g_{-n}\sigma_{g}^{2}(G_{n,n}-G_{n,-n}G_{-n,-n}^{-1}G_{-n,n})),\]
where positive/negative indices indicate the extraction/removal of
rows or columns. 

The second problem is the problem of sampling from $g_{-n}\mid y\;;\; G,\sigma_{g}^{2}$,
which is more involved. We introduce another set of variables, namely
the environmental effects $e_{-n}.$ It is then possible to write
down the conditional distribution of each variable in the set $(g_{-n},e_{-n})$, conditional on the rest of the variables in the set. 

The knowledge of the phenotype induces dependence between $g_{i}$
and $e_{i}$, since knowing the phenotype implies that we have an
upper or lower bound on their sum (the sum is either above or below
the threshold, depending on the phenotype). Additionally, $e_{i}$
is independent of the other environmental effects, and is also independent
of the other genetic effects conditional on $g_{i}$. Hence:
\[
e_{i}|g,e_{-i},y_{i};\sigma_{g}^{2}\;=\;e_{i}|g_{i},y_{i};\sigma_{g}^{2}\;=\;\begin{cases}
\sigma_{e}Z\mid\sigma_{e}Z+g_{i}>t\;\;\;\; & y_{i}=\text{case}\\
\sigma_{e}Z\mid\sigma_{e}Z+g_{i}<t & y_{i}=\text{control}\end{cases},\]
where $Z\sim N(0,1)$ (hence, the distribution is simply a truncated Normal). Intuitively, when $g_{i}$ is known and the
phenotype is known, the posterior distribution of the environmental
effect is a truncated Normal distribution, with mean $0$, variance
$\sigma_{e}^{2}$, and a truncation point above or below $t-g_{i}$,
depending on whether $i$ is a case or control.

The conditional distribution of $g_{i}$ is slightly more complicated,
as it depends on the other genetic effects via the correlation between
genetic effects. Denote $\mu_{i},\sigma_{i}^{2}$ the mean and variance
of $g_{i}$, conditional on $g_{-i}$, then:
\[
\mu_{i}=G_{i,-i}G_{-i,-i}^{-1}g_{-i},\]
and 
\[
\sigma_{i}^{2}=\sigma_{g}^{2}(G_{ii}-G_{i,-i}G_{-i,-i}^{-1}G_{-i,i}).\]

Similarly to the conditional distribution of the environmental effects,
conditioning on the phenotype results in a truncation of the aforementioned normal distribution:

\[
g_{i}|e_{i},g_{-i},y_{i};\sigma_{g}^{2}\;=\;\begin{cases}
\mu_{i}+\sigma_{i}Z\mid\mu_{i}+\sigma_{i}Z+e_{i}>t\;\;\;\; & y_{i}=\text{case}\\
\mu_{i}+\sigma_{i}Z\mid\mu_{i}+\sigma_{i}Z+e_{i}<t & y_{i}=\text{control}\end{cases}.\]

Again, the conditional distribution can be seen as a truncated normal
distribution, but with a mean term and standard deviation term which
capture the influence of the other genetic effects on the genetic
effect in question. 

Once all of the conditional distributions are specified, Gibbs sampling \cite{casella1992explaining} can be used to draw samples from the posterior distribution of $g_{-n}$, and the risk is estimated as described above. This is done is a similar fashion to \cite{campbell2010software}.\\ 

{\bf Simulation setup.}
Given the prevalence of a disease in the population ($K$), the desired proportion of cases in the study ($P$), the desired study size ($n$), the total number of SNPs ($m$), the proportion of slab SNPs ($\pi_1$), the overall variance of the genetic effects ($\sigma_g^2$) and the fraction of the heritability explained by the slab SNPs ($f_{slab}$, we simulated datasets using the following procedure:

\begin{enumerate}
\item The MAFs of $m$ SNPs were randomly sampled from $U[0.05,0.5]$.
\item SNP effect sizes for $\pi_{1}m$ slab SNPs, and $(1-\pi_{1})m$ spike SNPs, were randomly sampled
from $N(0,\frac{f_{slab}\sigma_{g}^{2}}{\pi_{1}m})$ and $N(0,\frac{(1-f_{slab})\sigma_{g}^{2}}{(1-\pi_{1})m})$, respectively.
\item For each individual, we:

\begin{enumerate}
\item Randomly generated a genotype using the MAFs.
\item Computed the genetic effect as described above. 
\item Sampled an environmental effect from $N(0,1-\sigma_{g}^{2})$.
\item Computed liability and phenotype.
\item If the phenotype was a case, the individual was automatically included in the study. Otherwise the individual was included in the study with probability $\frac{K(1-P)}{P(1-K)}$, to maintain the expected proportion of cases in the study at $P$.
\end{enumerate}
\item Step (2) was repeated until $n$ individuals were accumulated.
\end{enumerate}

Setting $f_{slab}=1$ results in a model where only $\pi_1$ of the SNPs are causal, and the rest of the SNPs have no effects on the phenotype.

We note that our choice to work with SNPs that are in linkage equilibrium
was motivated by a result of \cite{patterson2006population}. They show that for the purpose of generating correlation matrices, using SNPs in linkage disequilibrium (LD) is equivalent to using a smaller number of SNPs in linkage equilibrium. They also suggest a method for estimating
the effective number of SNPs (i.e. the number of SNPs in linkage equilibrium leading to the same distribution of correlation matrices as a given set of SNPs in LD). We thus find that our simulations using 50,000 SNPs in equilibrium are of realistic size.

{\bf Computing genetic risk scores using "standard" fixed effects models.}
To compute genetic risk scores using a fixed-effects approach, we follow the spirit of Dudbridge (2013) \cite{dudbridge2013power} and Chatterjee et al. (2013) \cite{chatterjee2013projecting}. For each SNP we estimate the effect size $\hat{u}$ using univariate linear regression. Denote by $v_{i}$ the p-value of the null hypothesis of $u_{i}=0$. We then define the estimated risk score using a p-value threshold $c$ as:
\[
\text{risk score}(c)_{j}=\sum\hat{u}_{i}Z_{ij}\mathbb{I}\{v_{i}<c\}.
\]
When dealing with real data, where LD structure is present, we select a subset of significant SNPs such that the distance between included SNPs is below 1Mb, by choosing the SNP with the lower p-value within any such window.
We note that other alternative definitions exist, e.g. using shrinkage estimates, but generally there's very little difference between the methods as noted by \cite{dudbridge2013power}.
For the real data, we try several p-value thresholds, namely $5\times10^{-c}$ for $c\in{1,...,8}$. We then choose the threshold which maximizes the AUC. Hence, AUC estimates of the fixed-effects model are expected to be slightly elevated. Our bootstrap scheme for computing confidence intervals accounts for this selection scheme as detailed in the Supplementary Materials. \\

{\bf Random and Mixed effects GeRSI.}
We use the genotyped SNPs to estimate the genetic correlation matrix $G$. When applying random-effects GeRSI, we use this matrix as the correlation matrix in the sampling scheme described above. 
When applying mixed-effects GeRSI with a a p-value threshold $c$, we keep only SNPs with a p-value below that threshold in the univariate association test. We then use logistic
regression to estimate the personal in-study risk due to the fixed effects. We then covert this risk to the liability scale using the following transformation:
\[
\hat{t}_{i}=\Phi(1-\frac{C\hat{P}_{i}}{1+C\hat{P}_{i}-\hat{P}_{i}}),
\]
where $\hat{P}_{i}$ is the estimated in-study risk, $\hat{t}_{i}$ is the individual-specific liability threshold and $C=\frac{K(1-P)}{P(1-K)}$ (for a full derivation, see \cite{golan2013narrowing}).\\ 

{\bf gBLUP.}
For comparison, we compute the genetic best linear unbiased predictor (gBLUP). The gBLUP is obtained by coding the discrete phenotype as 0/1 and treating is as a quantitative and randomly sampled phenotype.  In other words, the phenotype is modelled as:
\[
y\sim MVN(0,\Sigma),
\]
where $\Sigma=G\sigma_{g}^{2}+I\sigma_{e}^{2}$. Hence, if the phenotype
of individual $i$ is unknown, then the conditional mean of her phenotype
is easily given by the previous formula for the conditional mean of
a multivariate normal distribution:
\[
\mathbb{E}[y_{i}|y_{-i};\Sigma,\sigma_{g}^{2}]=\Sigma_{i,-i}\Sigma_{-i,-i}^{-1}y_{-i}.\]
The gBLUP method could be similarly extended to account for fixed effects (see, e.g., \cite{yang2011gcta}).\\

{\bf Controlling for population structure.} When attempting to control for population structure in association studies or heritability estimation, is is customary to include several top principal components as covariates. However, in the context of risk prediction, this would result in inflated estimates of the predictive accuracy. Instead, we remove the top $k$ principal components directly from the correlation matrix. In other words, denote $\lambda_1,...\lambda_n$ the sorted eigenvalues of the correlation matrix $G$, and denote $v_1,...,v_n$ their corresponding eigenvectors. We define a {\textit cleaned} correlation matrix as:
\[G^{*}(k) = G - \sum_{i=1}^{k}\lambda_i v_i v_i^\intercal.\]~\\ 

{\bf Real data description.}
We obtained genotypes and phenotypes from the WTCCC. 
Following Lee et al. (2011) \cite{lee2011estimating} we applied a stringent quality control (QC) process on the WTCCC data to avoid over-estimation of the predictive capacity due to genotyping differences between cases and controls or between the different control groups. We removed SNPs with MAF<5\%, with missing rate >1\% and SNPs which displayed a significantly different missing rate between cases and controls (p-value<0.05). We also removed SNPs which deviated from Hardy-Weinberg (HW) equilibrium in the control groups (p-value <0.05), or were noted for {}``bad clustering'' in the genotype calling step. Additionally we removed SNPs which displayed a significant difference in frequency between the two control groups. Only autosomal chromosomes were included in the analysis. We removed all the individuals appearing in the WTCCC exclusion lists. These include duplicate samples, first or second degree relatives, individuals which are not of European descent and other reasons. In addition we removed individuals with missing rate >1\% and all individual pairs with an estimated genetic correlation >0.05 based on the correlation matrix. The last step is done to ensure individuals in the study are not closely related. 
In addition, when computing the correlation matrix $G$, we used estimated minor-allele frequencies from HapMap's CEU panel \cite{gibbs2003international}, to mitigate any possibility of leakage between the train and test sets.\\

{\bf Implementation and availability.} Software implementing our methods is available from our website:\\ \url{https://sites.google.com/site/davidgolanshomepage/software/gersi}.

\bibliographystyle{unsrtnat}
\bibliography{references}
\end{document}


\title{Supplementary Materials for ``Effective Genetic Risk Prediction Using Mixed Models''}

\author{David Golan and Saharon Rosset }

\maketitle
\section*{Gibbs sampling in case-control studies}

The fact that the observed samples are obtained via a case-control
sampling scheme, and are therefore not a random sample from the population,
renders the usual mixed-effects model incompatible. In particular,
under assumptions of (a) normality of genetic and environmental effects
in the population and (b) Independence of the genetic and environmental
effects in the population, the actual distribution of genetic and
environmental effects in the study is non-normal and they are not
independent, due to selection, as noted in \cite{lee2011estimating}. Since the
distribution of the genetic effects is not longer normal, their joint
distribution is no longer multi-variate normal, and so, naive application
of the Gibbs sampling might be inaccurate. However, we show here that
the same sampling scheme can be used to sample the posterior genetic
effects in a case-control study. 

To model and account for the effects of selection, we define an event
${\cal S}$ which signifies that individuals $1,...,n-1$ were selected
via a case-control scheme and not by random sampling. Hence, the conditional
distributions above now require an additional conditioning on ${\cal S}$.
However, we note that:

\[
f(g_{n}|g_{-n},{\cal S};G,\sigma_{g}^{2})=\frac{f(S|g_{-n},g_{n};G,\sigma_{g}^{2})f(g_{n}\mid g_{-n};G,\sigma_{g}^{2})}{f(S|g_{-n};G,\sigma_{g}^{2})},\]

but since ${\cal S}$ signifies only the selection of individuals $1$
through $n-1$, it is independent of $g_{n}$, hence $f(S|g_{-n},g_{n};G,\sigma_{g}^{2})=f(S|g_{-n};G,\sigma_{g}^{2})$
and so: \[
f(g_{n}|g_{-n},{\cal S};G,\sigma_{g}^{2})=f(g_{n}|g_{-n};G,\sigma_{g}^{2}),\]
i.e. Gibbs sampling of the genetic effect of the individual in question
can be carried out as if there's no selection, given that the samples
of the genetic effects $g_{1},...,g_{n-1}$ are drawn by correctly
conditioning on ${\cal S}$. Moreover, for an individual $i$ in the
reference group, we have:
\[
f(g_{i}|e_{i},g_{-i},y_{i},{\cal S};\sigma_{g}^{2})=\frac{f({\cal S}|e_{i},g_{-i},y_{i},g_{i};\sigma_{g}^{2})f(g_{i}|e_{i},g_{-i},y_{i};\sigma_{g}^{2})}{f({\cal S}|e_{i},g_{-i},y_{i};\sigma_{g}^{2})}\]
but since we assumed that the selection is driven only by the phenotypes,
we have $f({\cal S}|e_{i},g_{-i},y_{i};\sigma_{g}^{2})=f({\cal S}|y_{i};\sigma_{g}^{2})$
and $f({\cal S}|e_{i},g_{-i},y_{i},g_{i};\sigma_{g}^{2})=f({\cal S}|y_{i};\sigma_{g}^{2})$
so again the sampling boils down to the same Gibbs scheme. Lastly,
we need take care of the sampling of the environmental effects, but
since this is done per individual, the selection has no effect. To
conclude, the same Gibbs sampling scheme can be applied to case-control
studies, and yield correct posterior risk estimates.

\section*{Additional Simulation results}

\subsection*{Using the simple causative model}

\begin{figure}[H]
\centering
\includegraphics[scale=0.5]{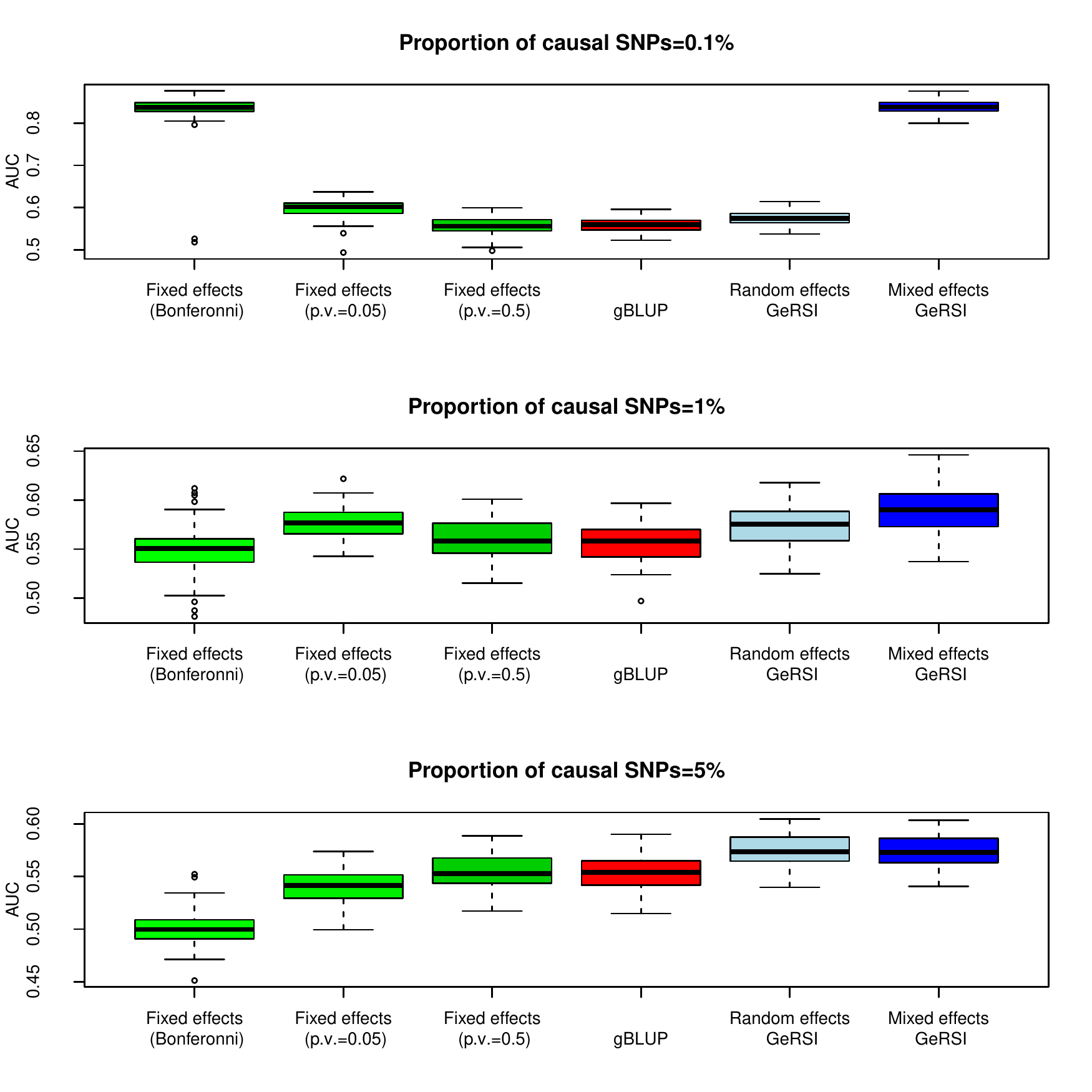}
\caption{Simulating effects using the normal distribution. K=0.05.}
\label{norm_1}
\end{figure}

\begin{figure}[H]
\centering
\includegraphics[scale=0.5]{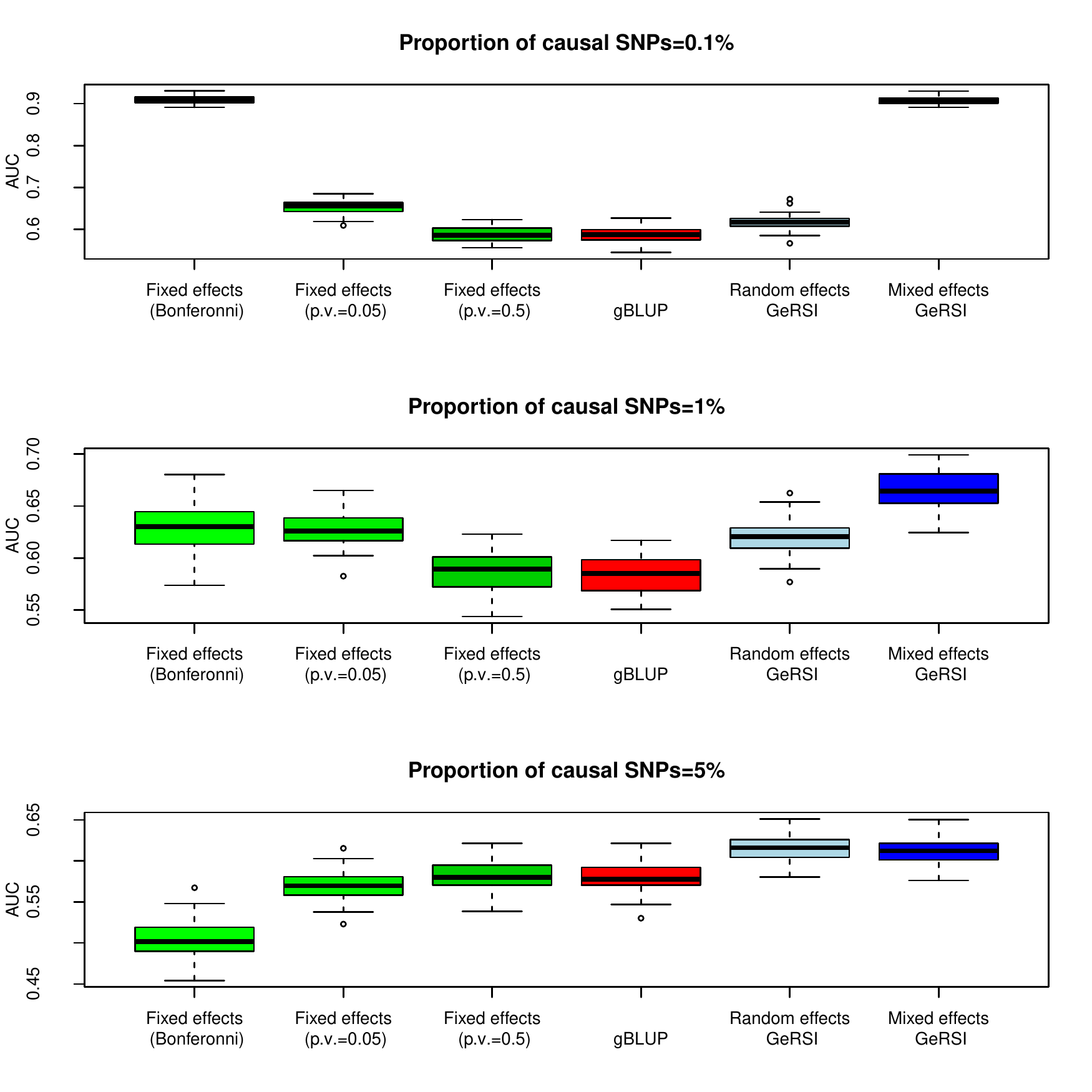}
\caption{Simulating effects using the normal distribution. K=0.01.}
\label{norm_2}
\end{figure}

\begin{figure}[H]
\centering
\includegraphics[scale=0.5]{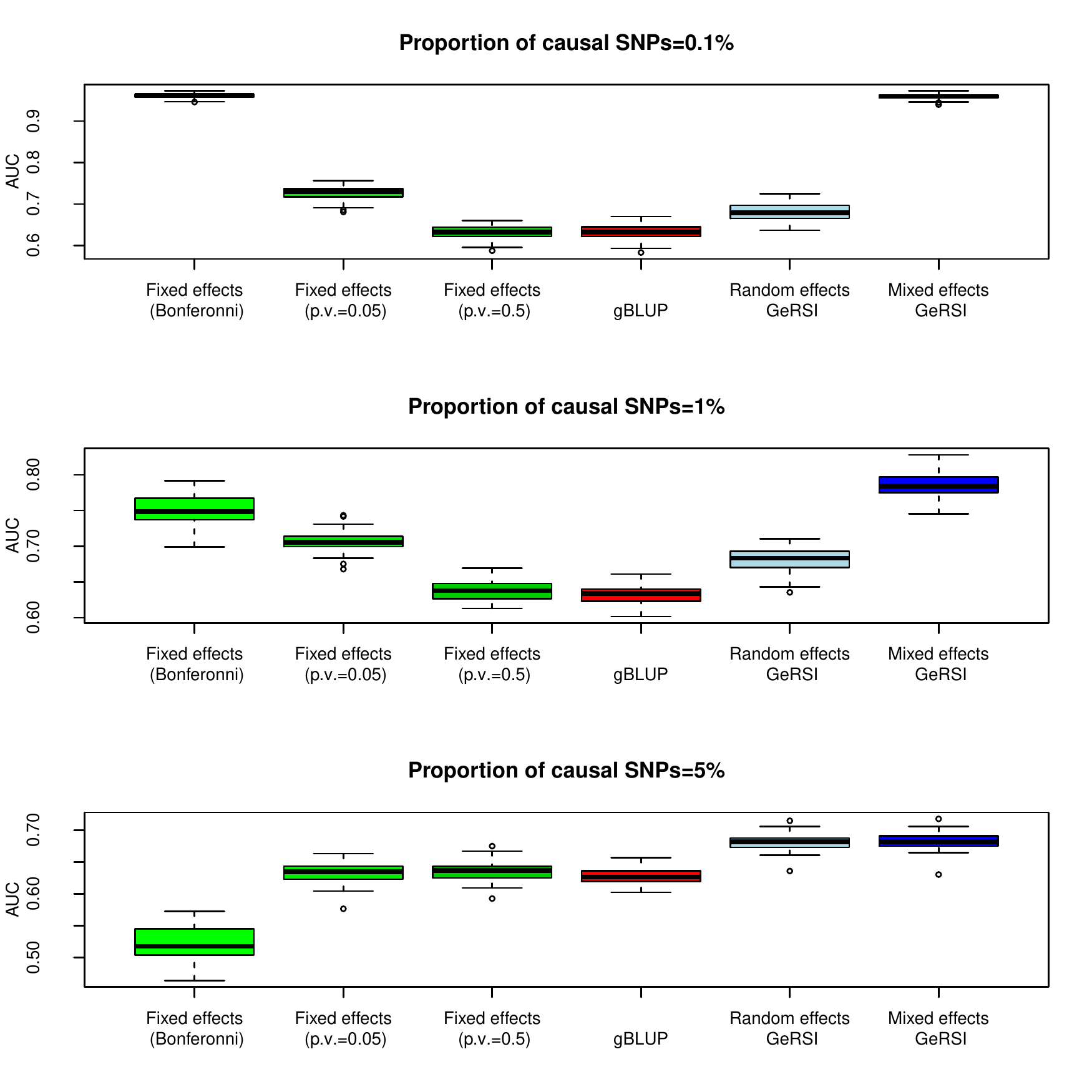}
\caption{Simulating effects using the normal distribution. K=0.001.}
\label{norm_3}
\end{figure}

We explored the performance of the various methods when a fraction $\pi_1$ of the SNPs are causal, while the rest of the SNPs have no effects. This allows for simulations with a larger number of SNPs as only the causative SNPs need to be simulated for the entire population, and the non-causal SNPs can be simulated only for the individuals selected for the study. The results are given in figures \ref{norm_1}-\ref{norm_3}.

\subsection*{Using the double-exponential distribution to model effect sizes}

\begin{figure}[H]
\centering
\includegraphics[scale=0.5]{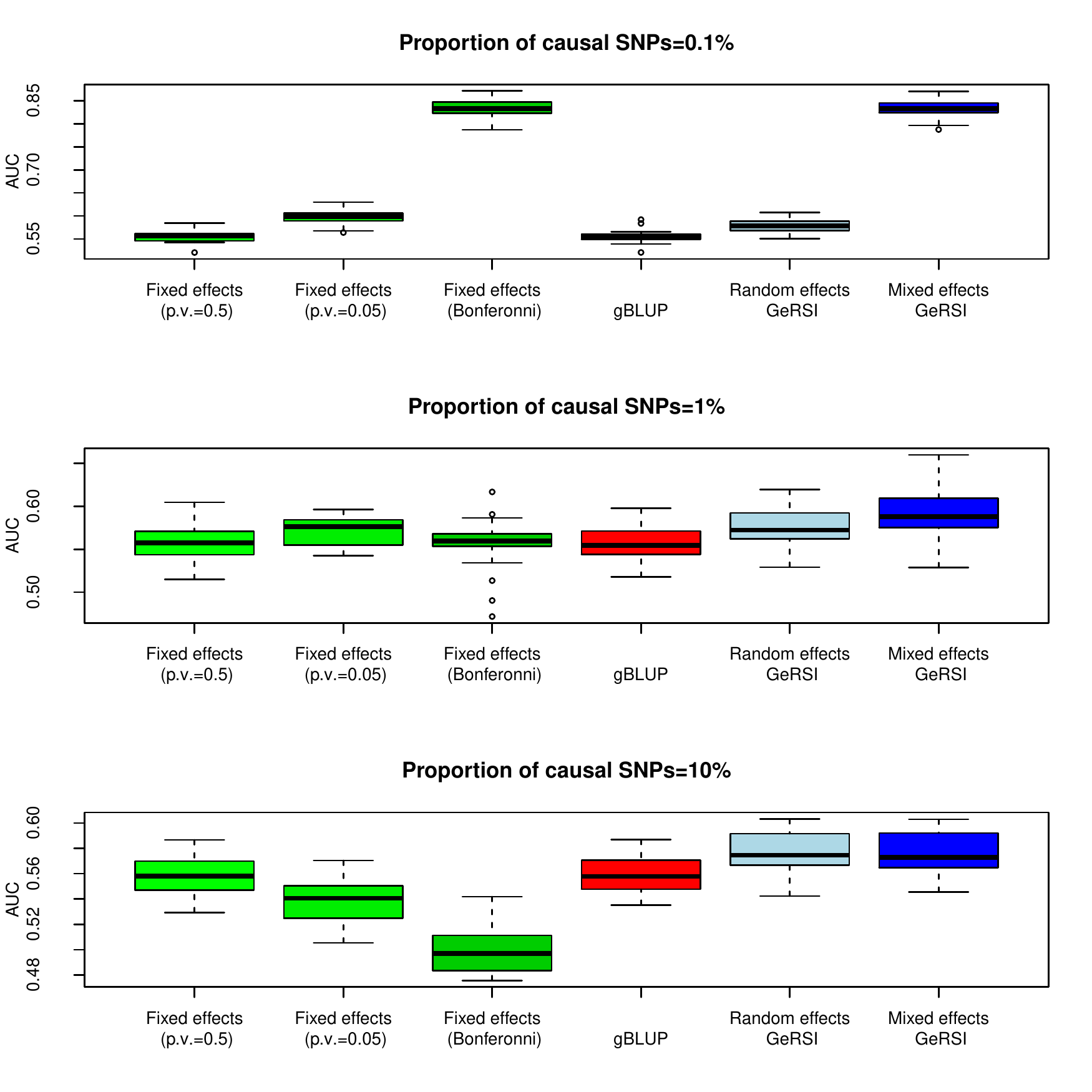}
\caption{Simulating effects using the double exponential distribution. K=0.05.}
\label{dexp_1}
\end{figure}

\begin{figure}[H]
\centering
\includegraphics[scale=0.5]{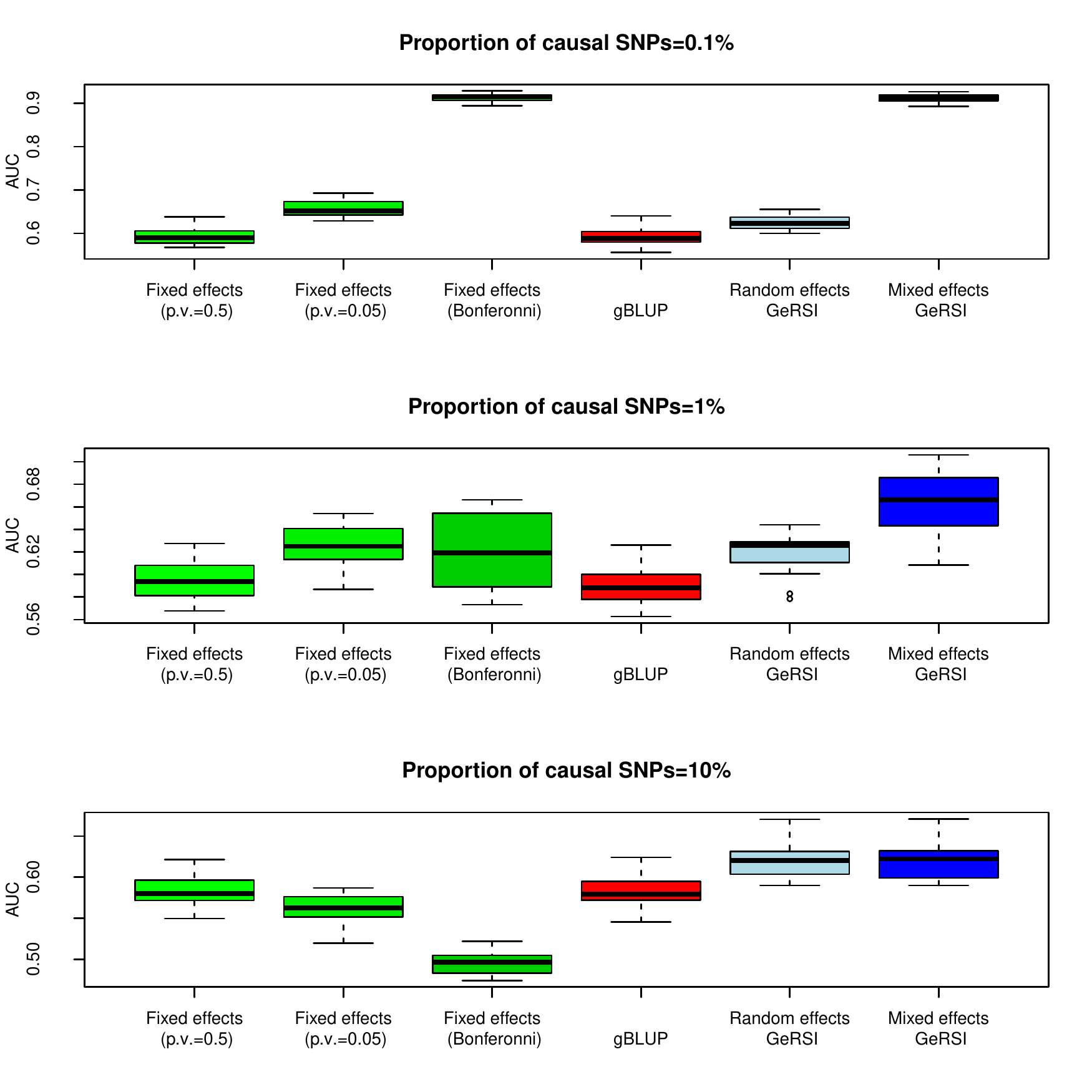}
\caption{Simulating effects using the double exponential distribution. K=0.01.}
\label{dexp_2}
\end{figure}

\begin{figure}[H]
\centering
\includegraphics[scale=0.5]{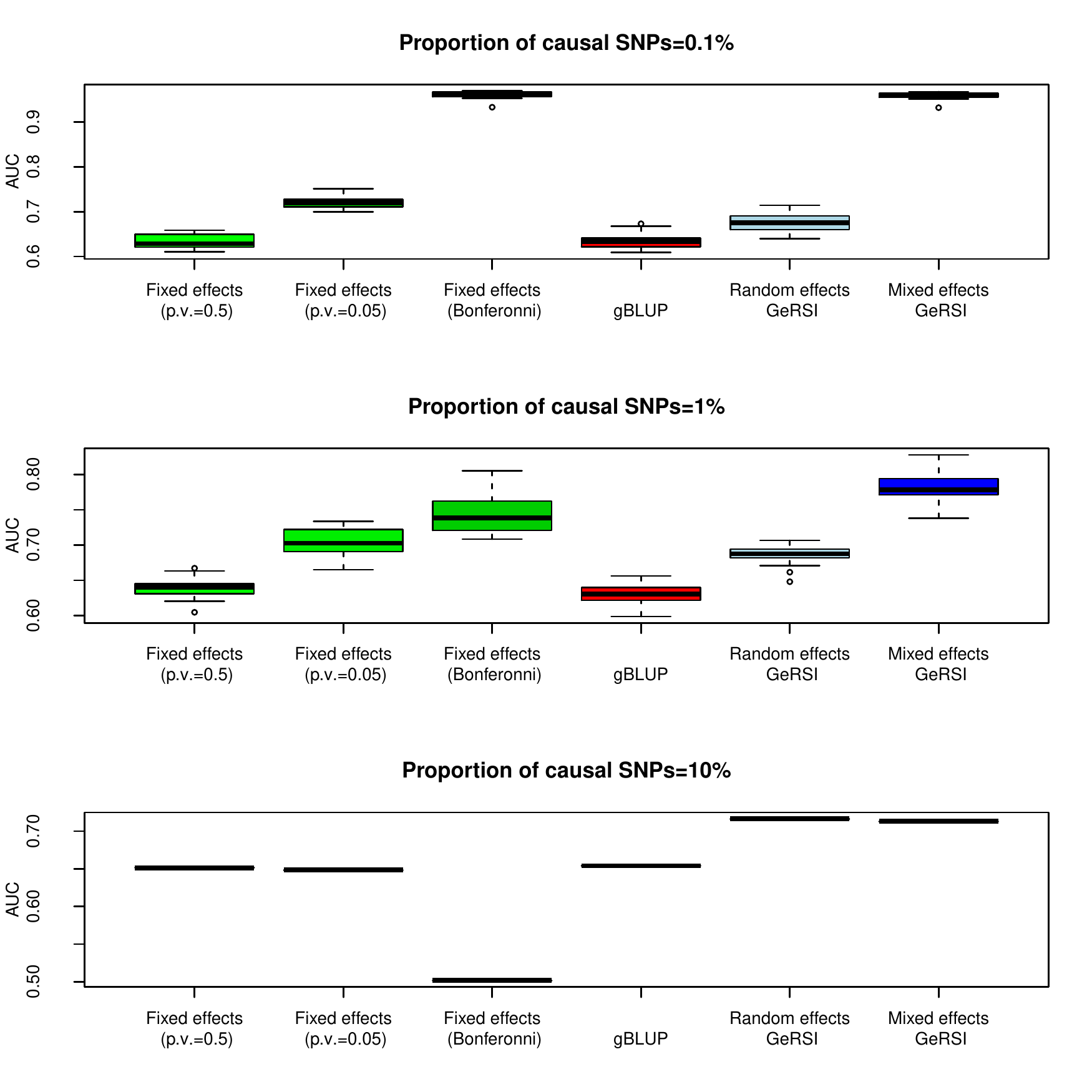}
\caption{Simulating effects using the double exponential distribution. K=0.001.}
\label{dexp_3}
\end{figure}

To study robustness of GeRSI to assumptions about the distribution
of effect sizes, we reran the same simulation scheme, but with the
effect sizes drawn from a double-exponential distribution with the
parameter calibrated to achieve the appropriate variance. The results
are given in figures \ref{dexp_1}-\ref{dexp_3}, but qualitatively the results are the same.

\subsection*{Using mixture distributions to model effect sizes}

Chatterjee et al. (2013) \cite{chatterjee2013projecting} and Zhou et al. (2013) \cite{zhou2013polygenic} argue that mixture models provide
a more realistic model of the distribution of effect sizes of SNPs.
To study the effect of such {}``spike and slab'' models on the performance
of GeRSI and other risk prediction methods, we draw the effect sizes
from the following distribution:

\[
u\sim\pi_{1}N(0,\sigma_{slab}^{2})+(1-\pi_{1})N(0,\sigma_{spike}^{2}).\]

\begin{figure}[H]
\centering
\includegraphics[scale=0.5]{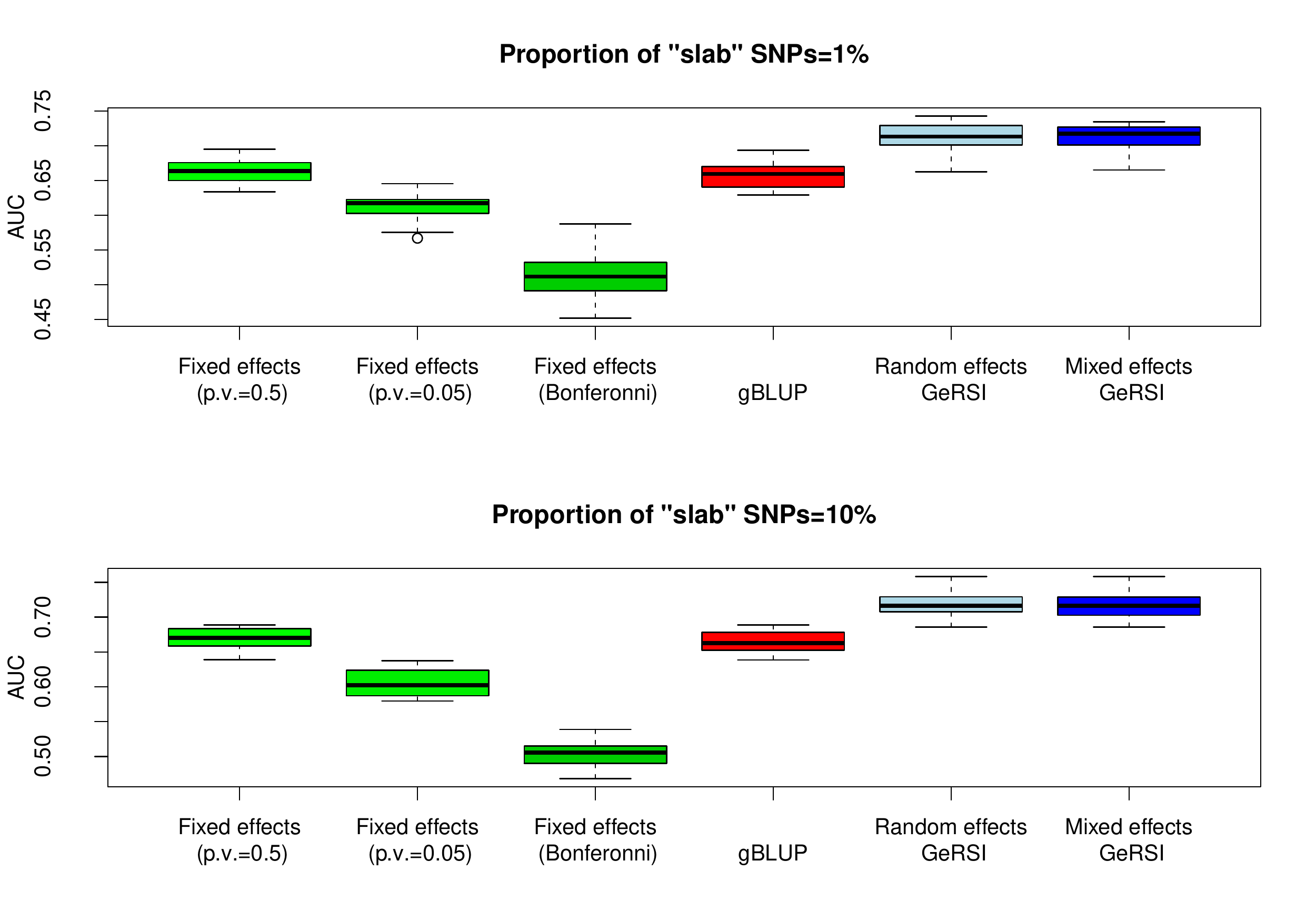}
\caption{K=0.05, proportion of heritability due to slab=0.1.}
\label{mix_1}
\end{figure}

\begin{figure}[H]
\centering
\includegraphics[scale=0.5]{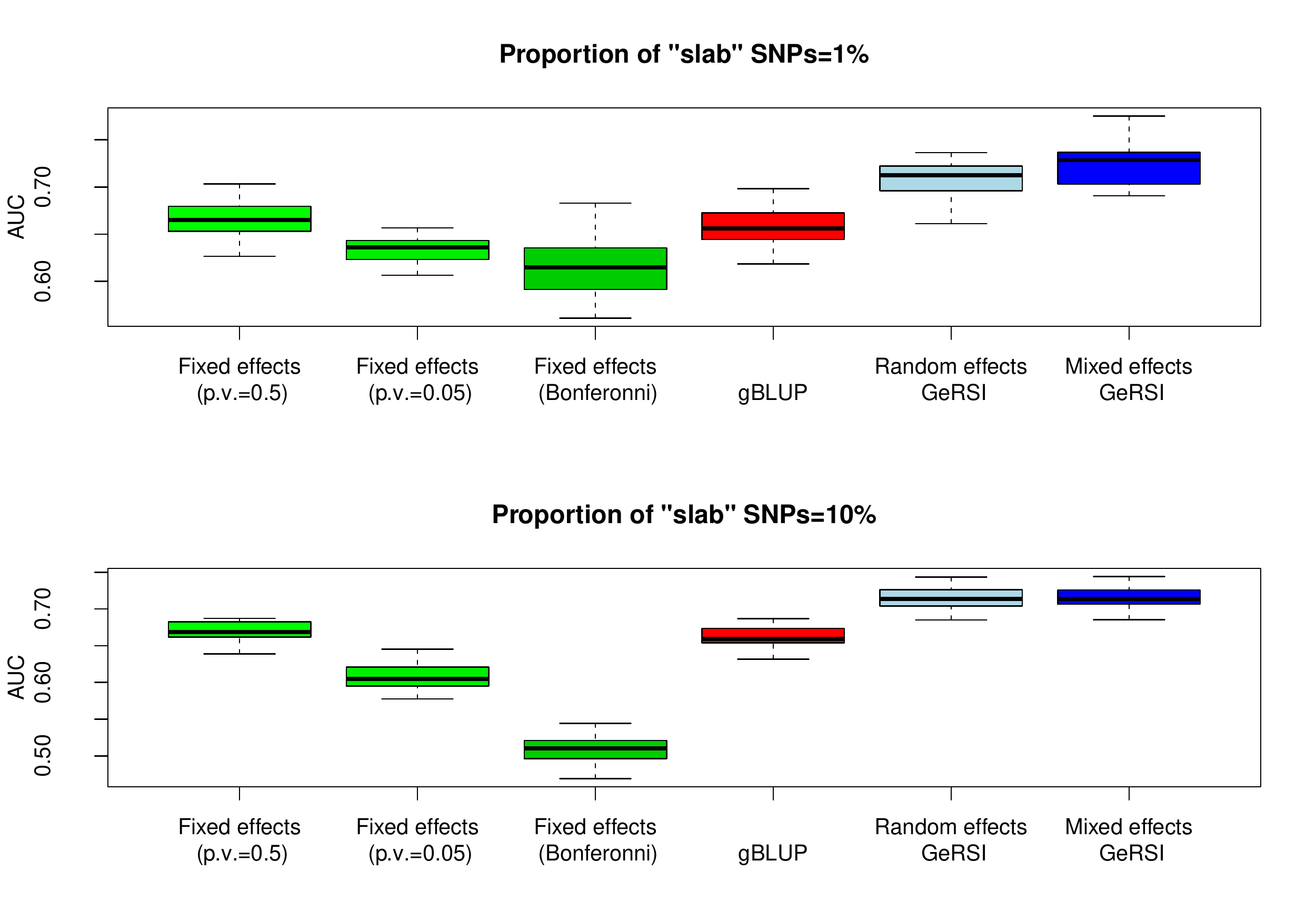}
\caption{K=0.05, proportion of heritability due to slab=0.3.}
\label{mix_2}
\end{figure}

\begin{figure}[H]
\centering
\includegraphics[scale=0.5]{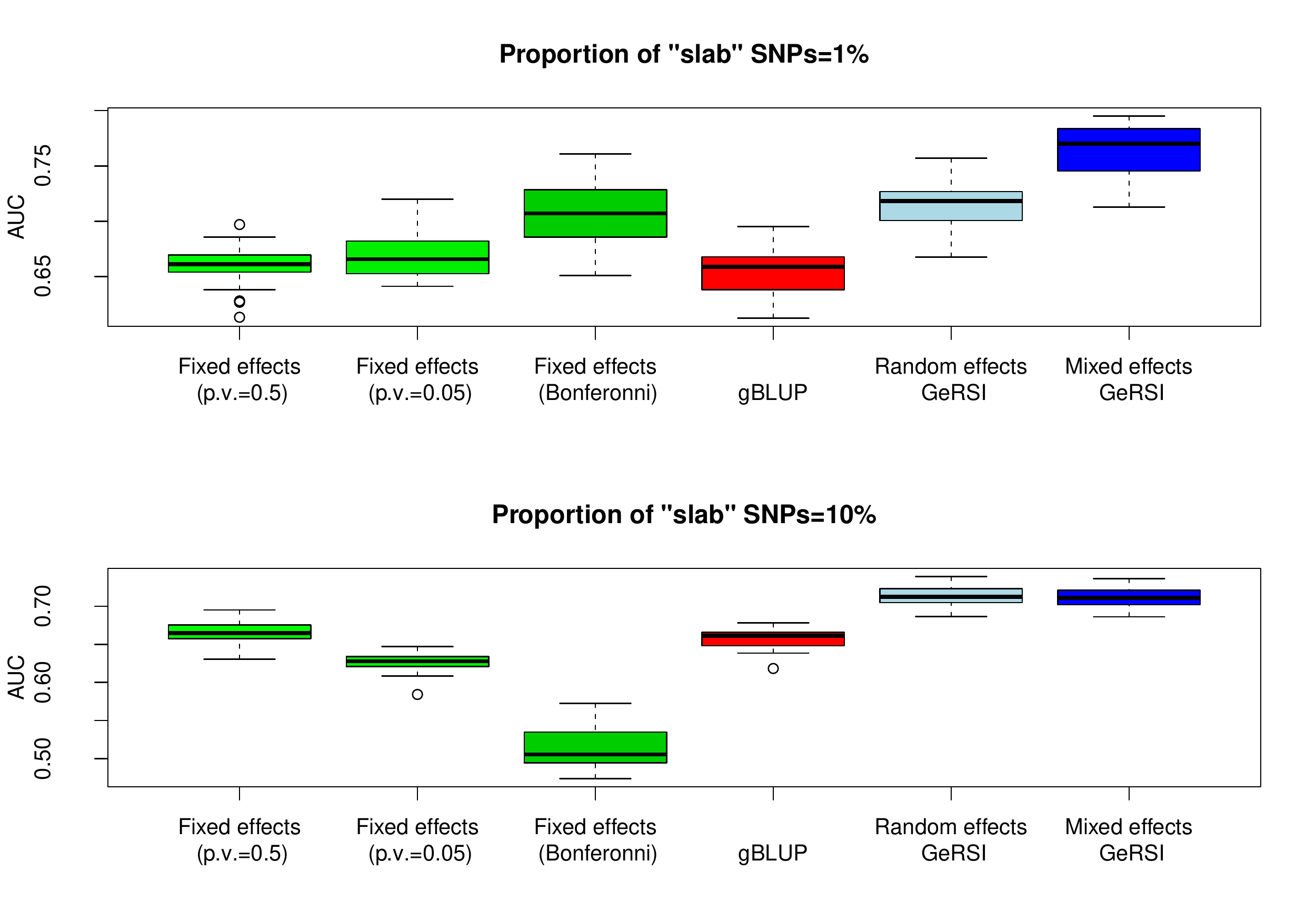}
\caption{K=0.05, proportion of heritability due to slab=0.5.}
\label{mix_3}
\end{figure}

\begin{figure}[H]
\centering
\includegraphics[scale=0.5]{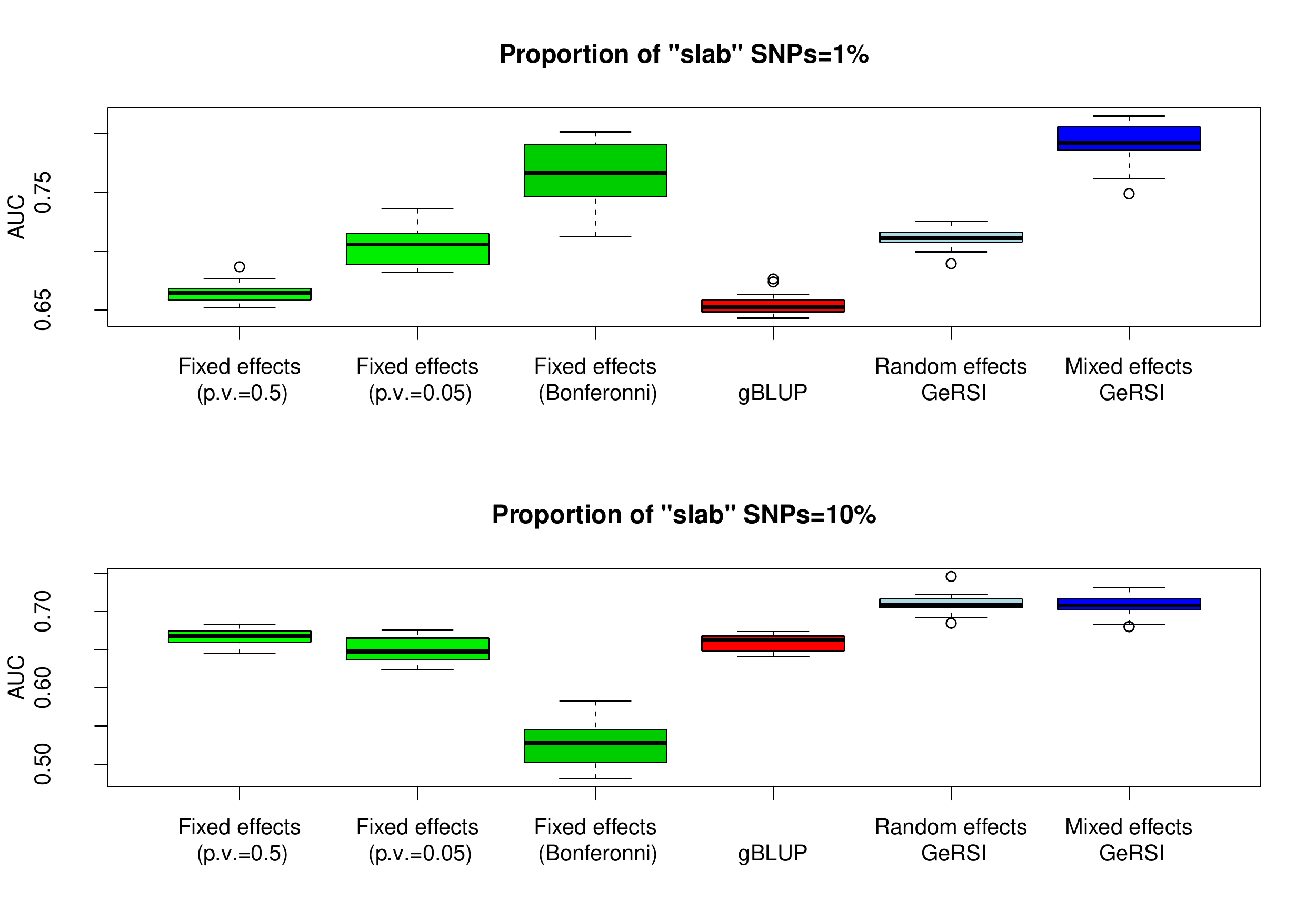}
\caption{K=0.05, proportion of heritability due to slab=0.7.}
\label{mix_4}
\end{figure}

\begin{figure}[H]
\centering
\includegraphics[scale=0.5]{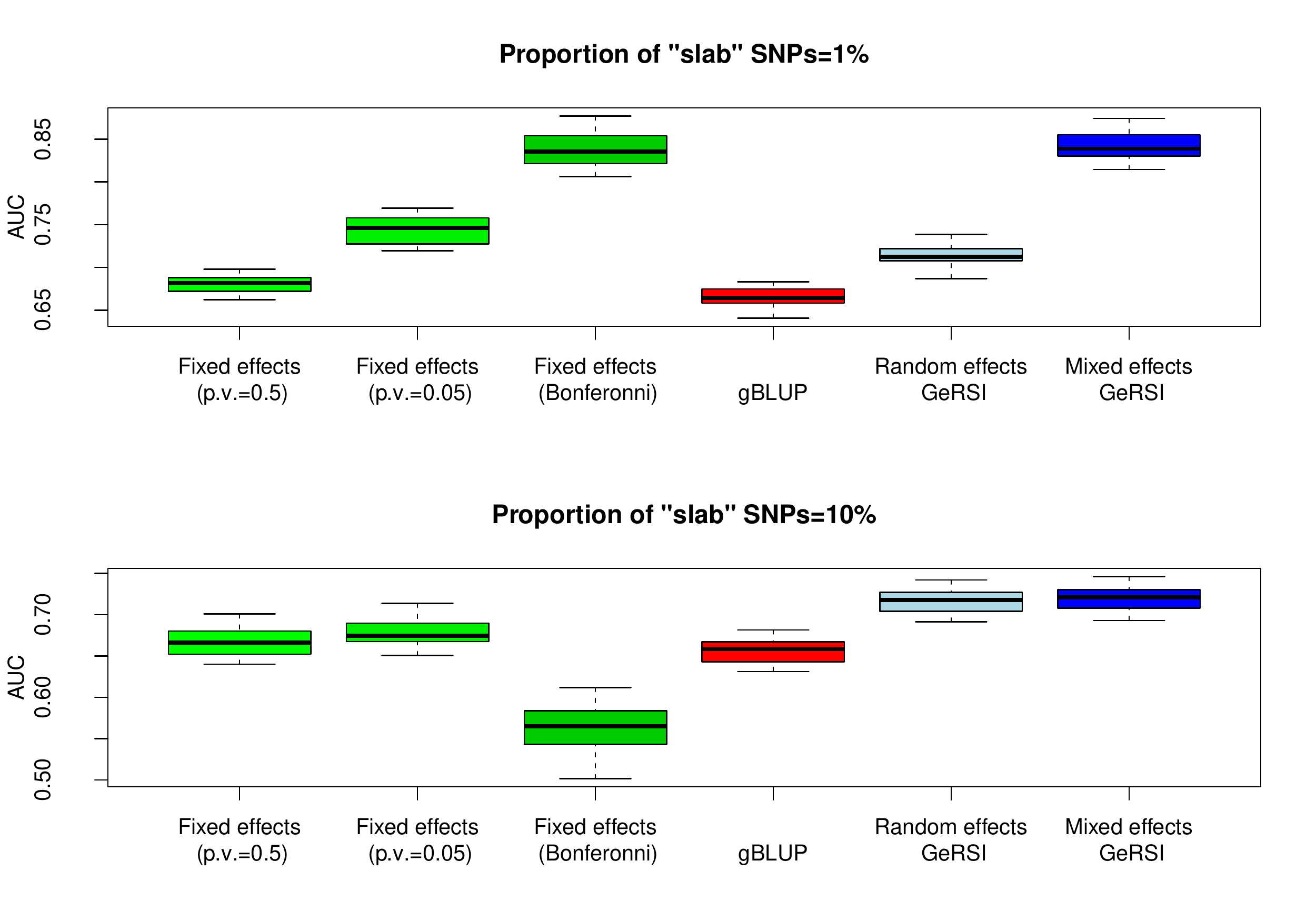}
\caption{K=0.05, proportion of heritability due to slab=0.9.}
\label{mix_5}
\end{figure}

\begin{figure}[H]
\centering
\includegraphics[scale=0.5]{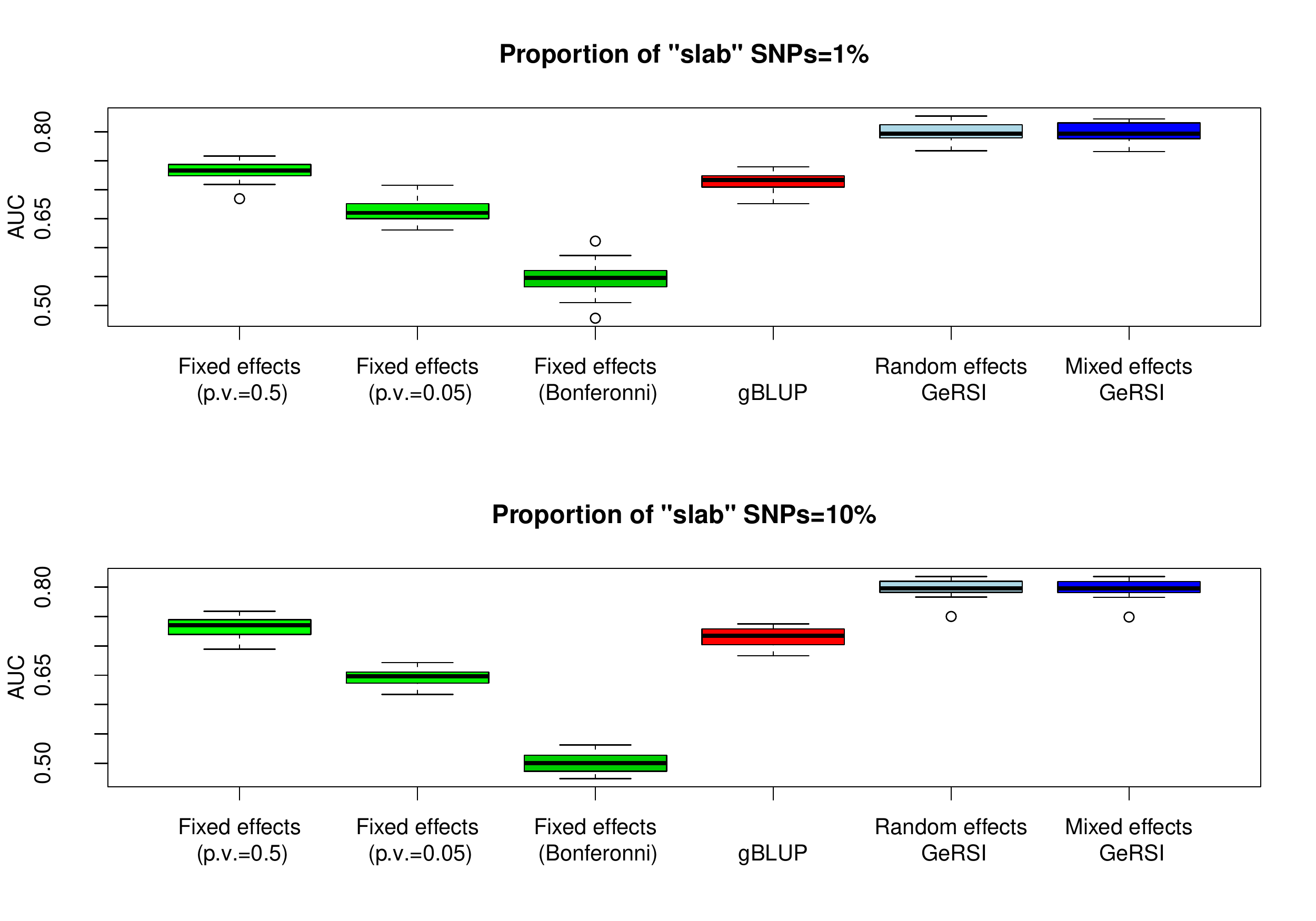}
\caption{K=0.01, proportion of heritability due to slab=0.1.}
\label{mix_6}
\end{figure}

\begin{figure}[H]
\centering
\includegraphics[scale=0.5]{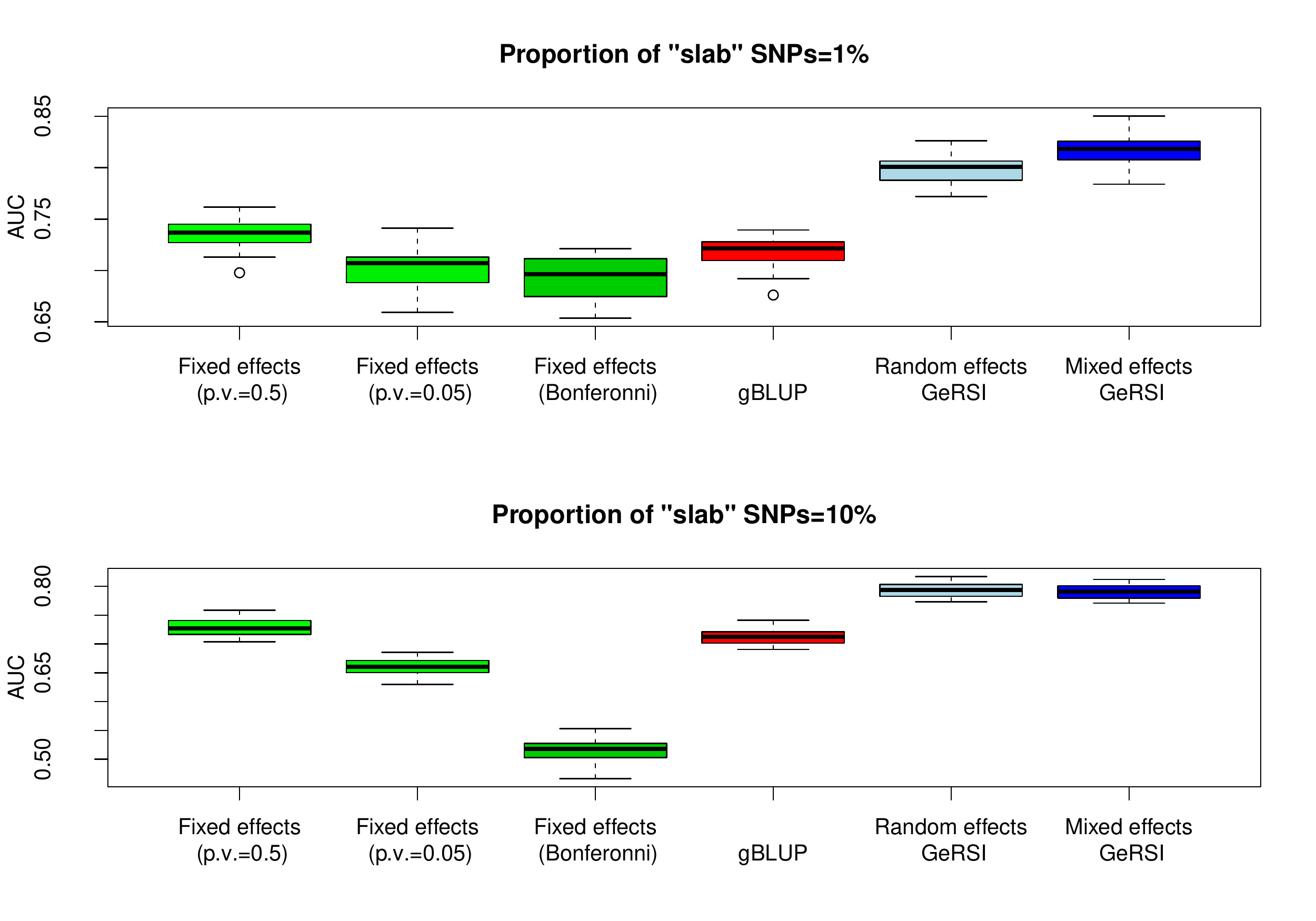}
\caption{K=0.01, proportion of heritability due to slab=0.3.}
\label{mix_7}
\end{figure}

\begin{figure}[H]
\centering
\includegraphics[scale=0.5]{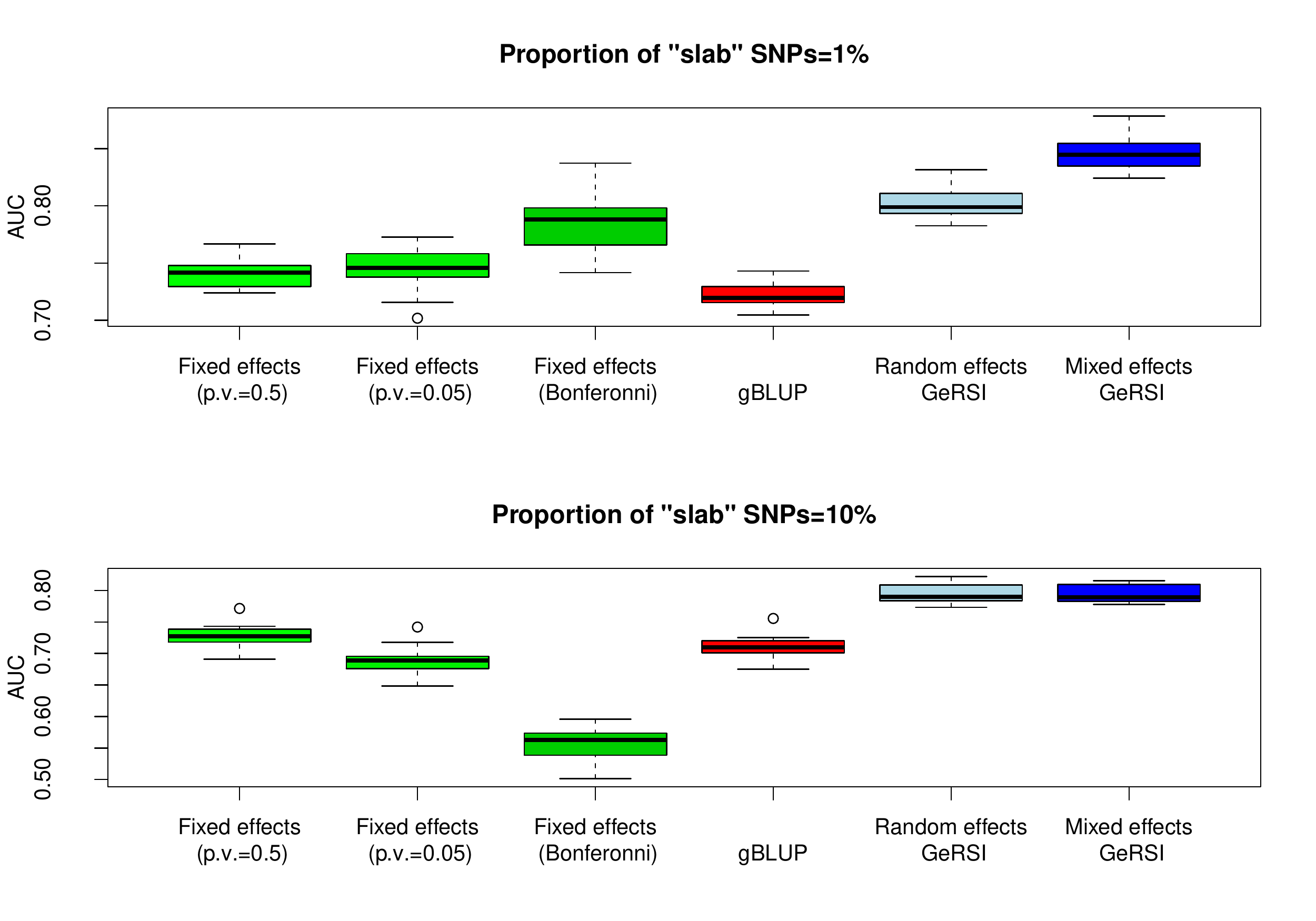}
\caption{K=0.01, proportion of heritability due to slab=0.5.}
\label{mix_8}
\end{figure}

\begin{figure}[H]
\centering
\includegraphics[scale=0.5]{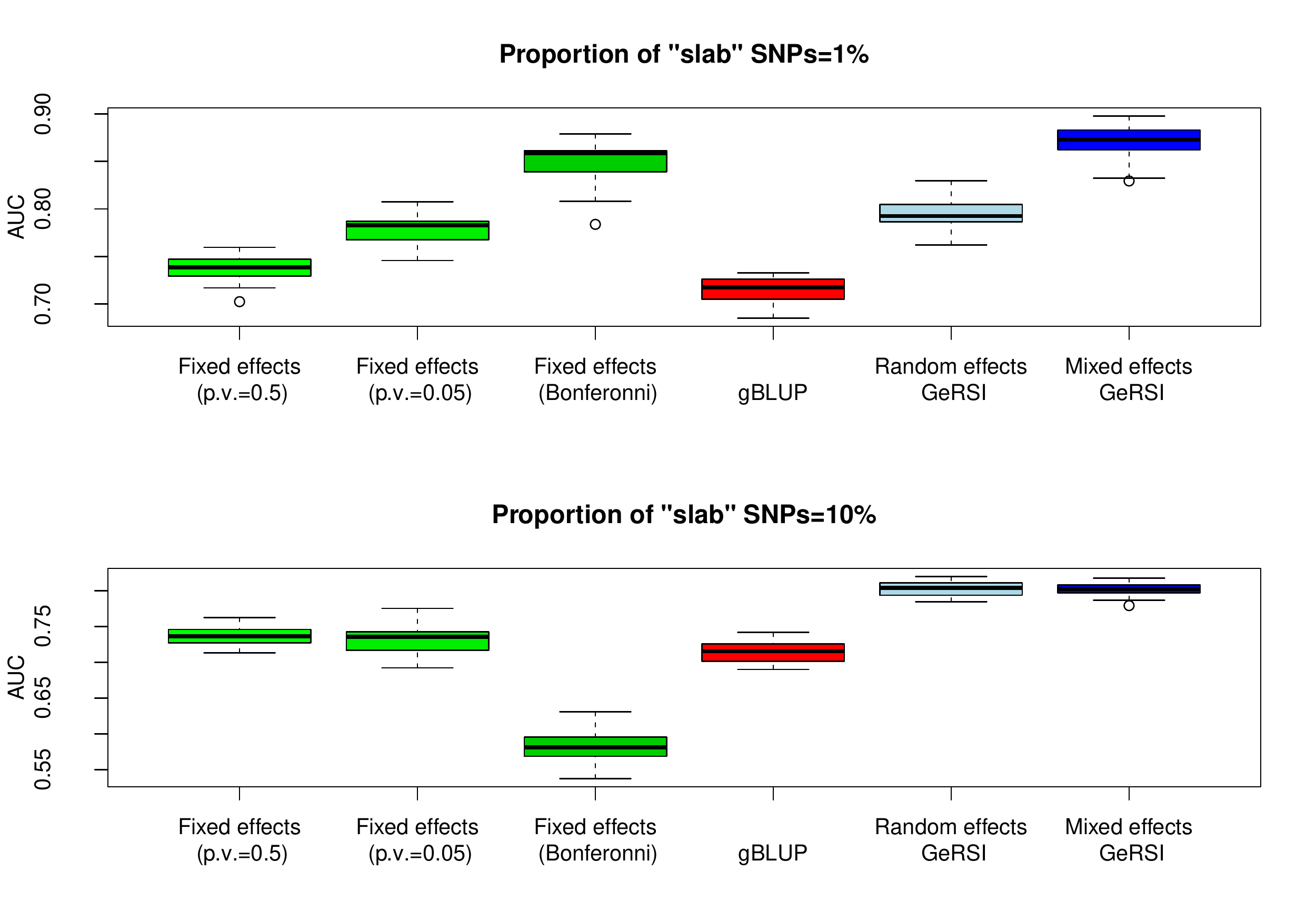}
\caption{K=0.01, proportion of heritability due to slab=0.7.}
\label{mix_9}
\end{figure}

\begin{figure}[H]
\centering
\includegraphics[scale=0.5]{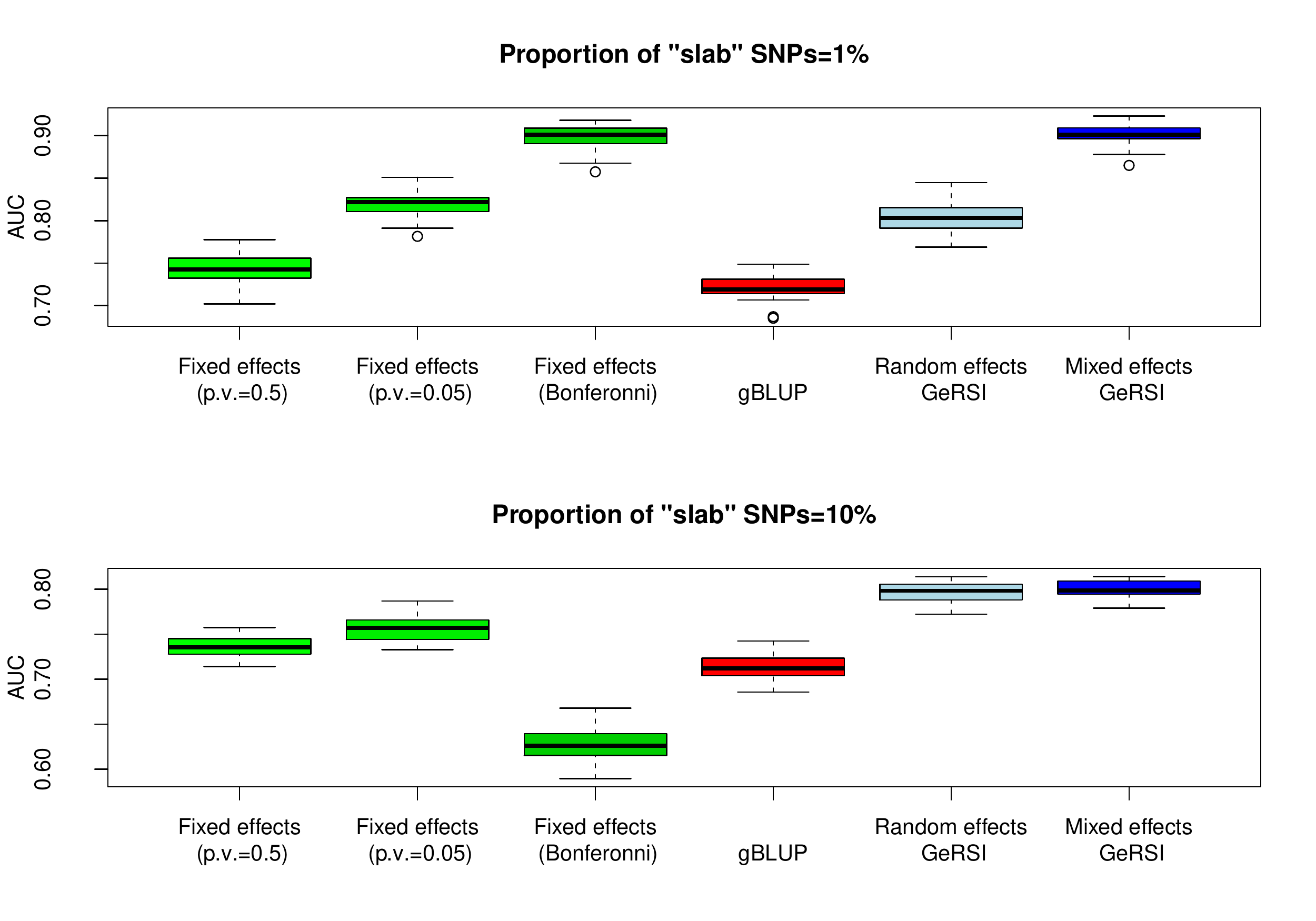}
\caption{K=0.01, proportion of heritability due to slab=0.9.}
\label{mix_10}
\end{figure}

\begin{figure}[H]
\centering
\includegraphics[scale=0.5]{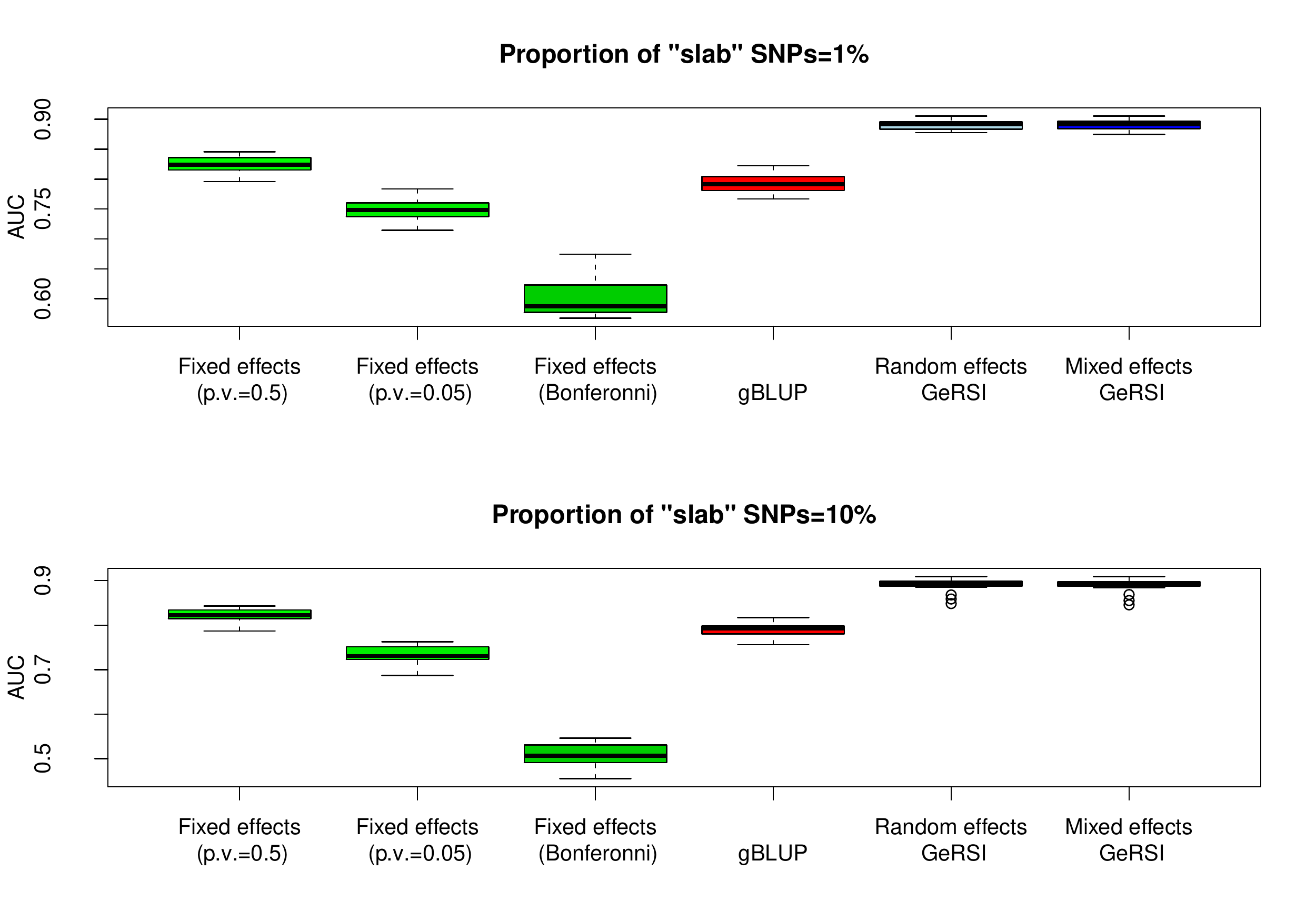}
\caption{K=0.001, proportion of heritability due to slab=0.1.}
\label{mix_11}
\end{figure}

\begin{figure}[H]
\centering
\includegraphics[scale=0.5]{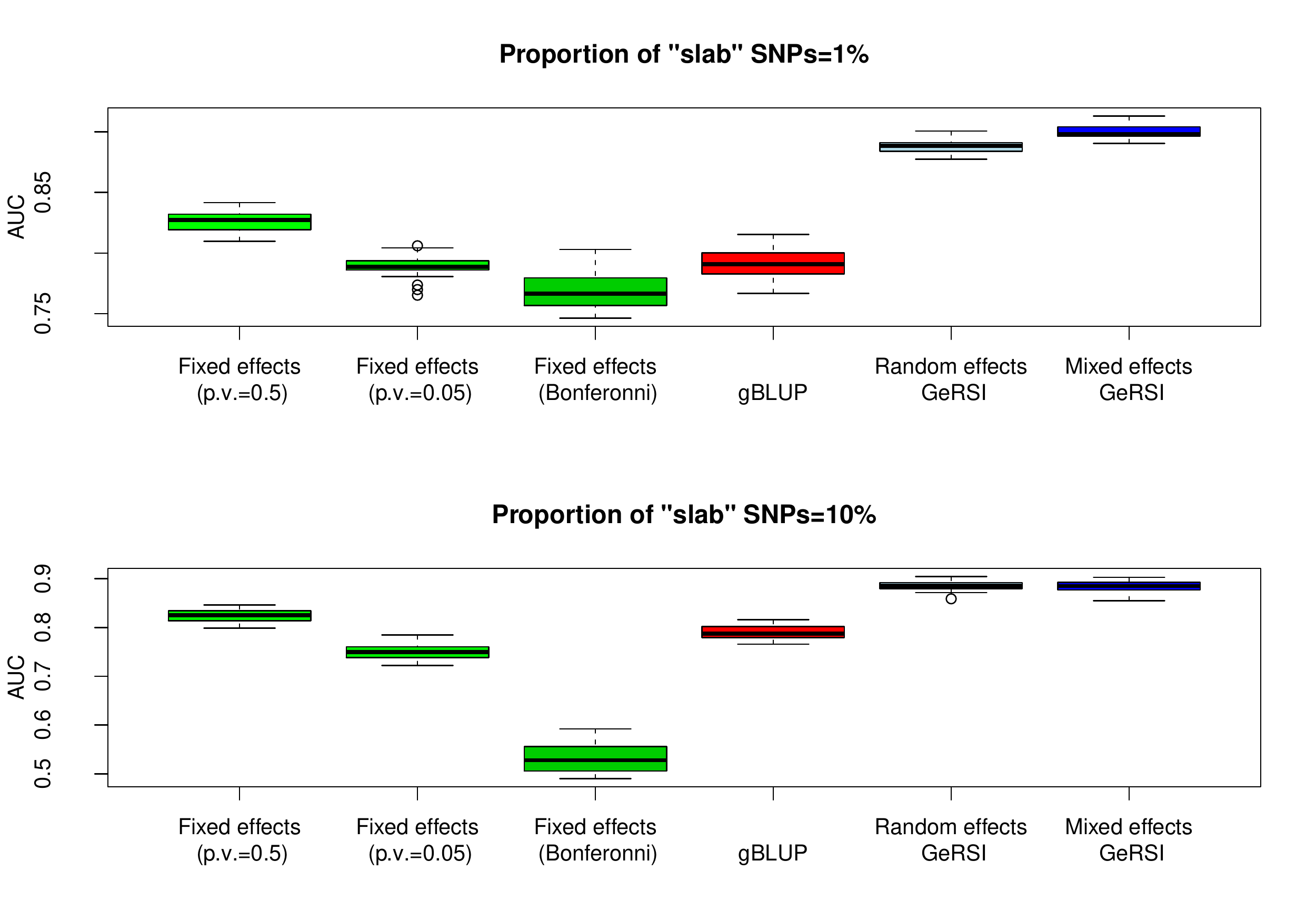}
\caption{K=0.001, proportion of heritability due to slab=0.3.}
\label{mix_12}
\end{figure}

\begin{figure}[H]
\centering
\includegraphics[scale=0.5]{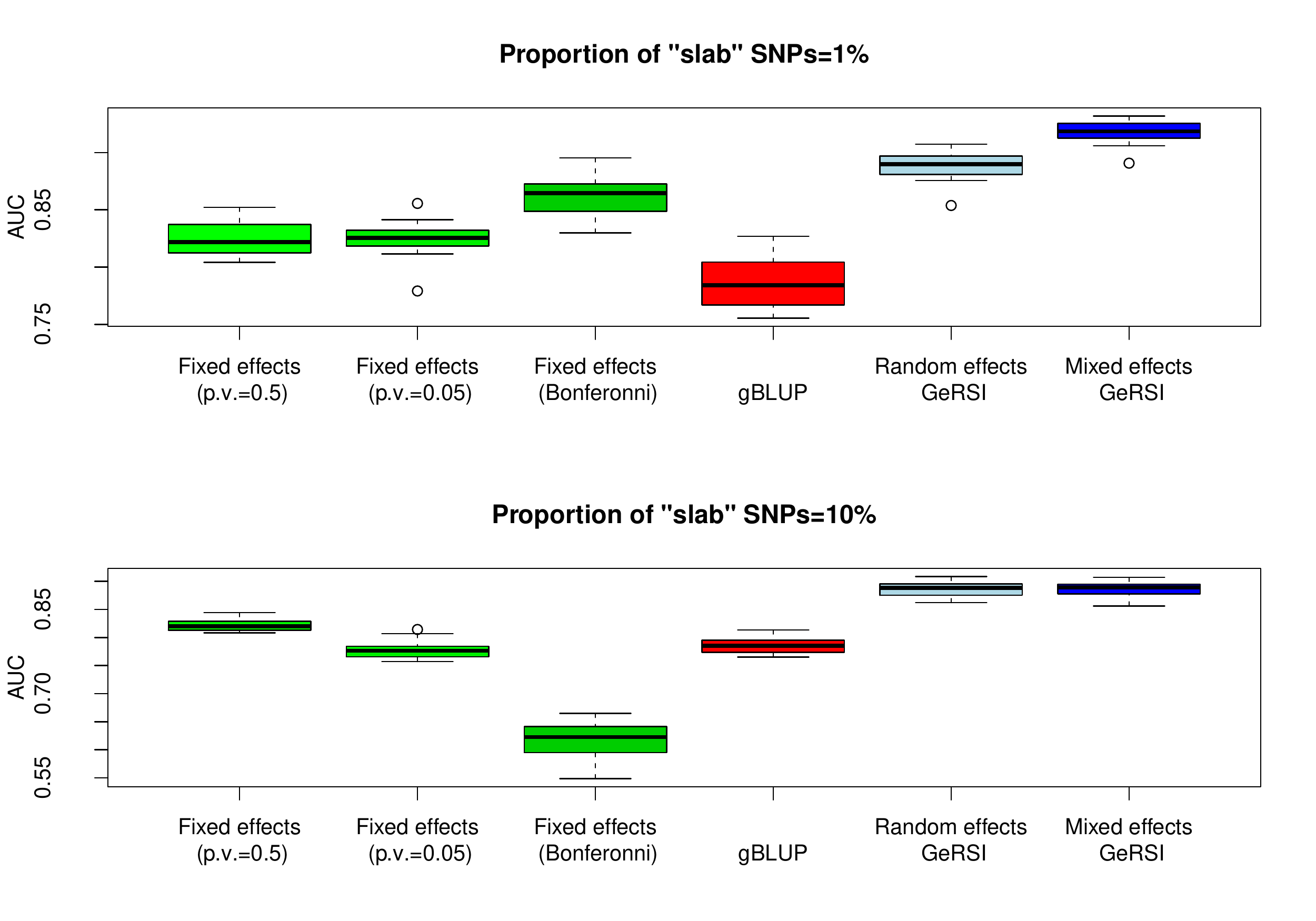}
\caption{K=0.001, proportion of heritability due to slab=0.5.}
\label{mix_13}
\end{figure}

\begin{figure}[H]
\centering
\includegraphics[scale=0.5]{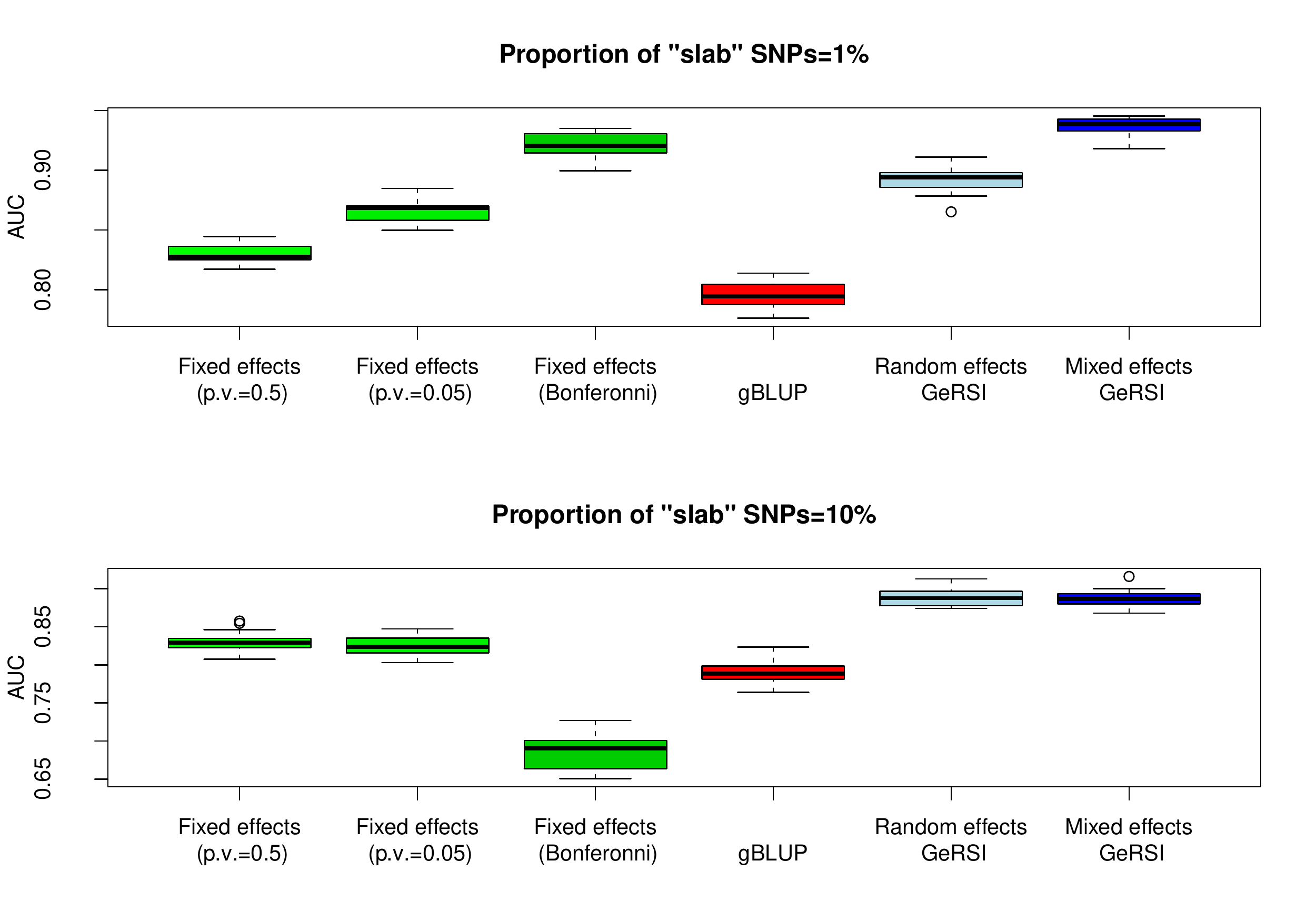}
\caption{K=0.001, proportion of heritability due to slab=0.7.}
\label{mix_14}
\end{figure}

\begin{figure}[H]
\centering
\includegraphics[scale=0.5]{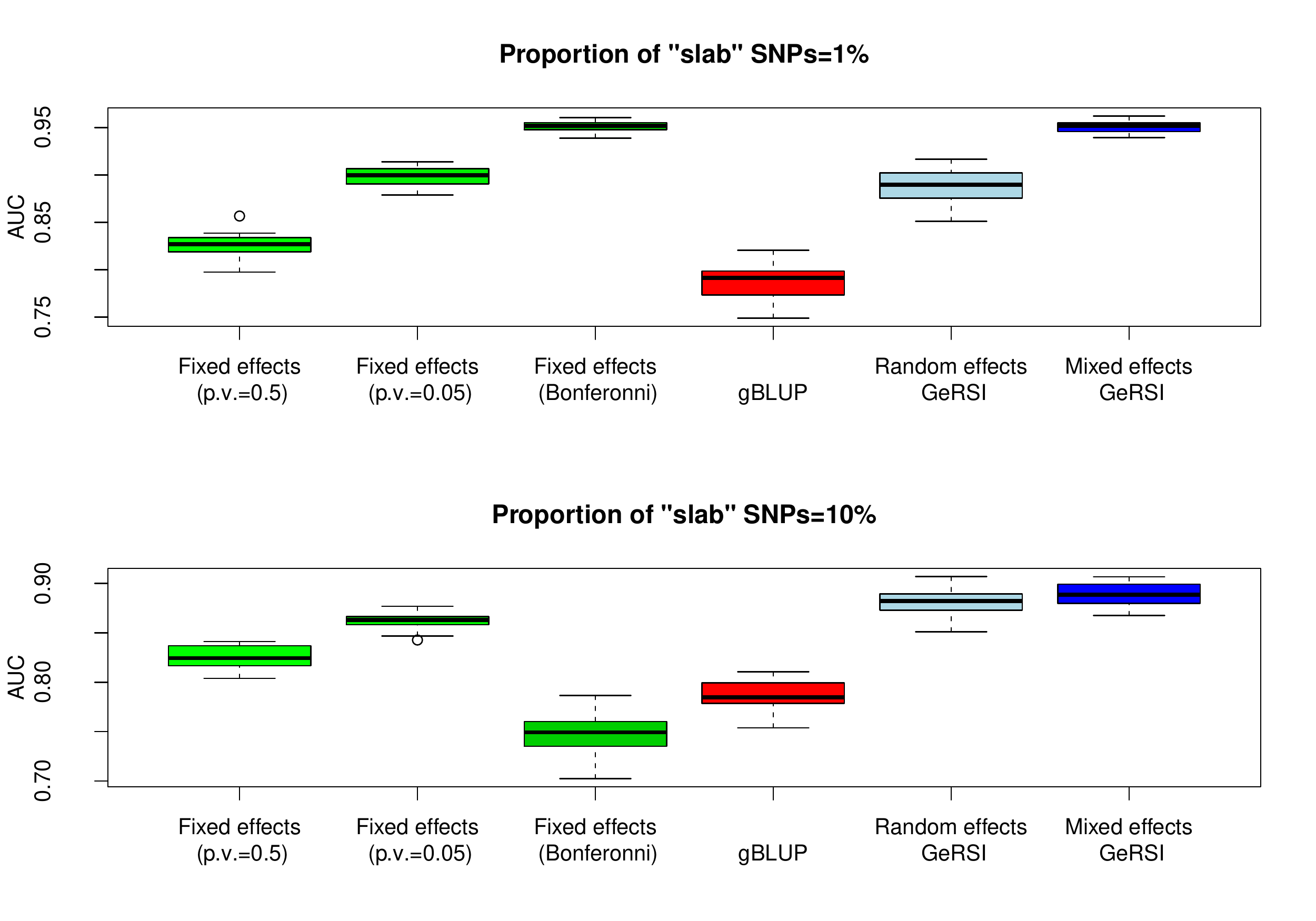}
\caption{K=0.001, proportion of heritability due to slab=0.9.}
\label{mix_15}
\end{figure}

Note that $\sigma_{u}^{2}=\pi_{1}\sigma_{slab}^{2}+(1-\pi_{1})\sigma_{spike}^{2}$.
We simulate genotypes and phenotypes from this model for all combinations
of $\pi_{1}\in\{0.1\%,1\%,10\%\}$, and $\frac{\sigma_{spike}^{2}}{h^{2}}\in\{10\%,30\%,50\%,70\%,90\%\}$,
while fixing the heritability at $50\%$. The results of these simulations
are shown in figures \ref{mix_1}-\ref{mix_15}.

We note that Chatterjee et al. (2013) estimate the components of such mixtures
for a wide range of phenotypes. resulting in a wide range of values.
We chose the values of our simulations to cover a considerable part
of the parameters estimated by Chatterjee et al. (2013).

As might be expected intuitively, under a mixture model, mixed-effect
GeRSI considerably outperforms all other approaches.

\section*{Inference}
We use the bootstrap \cite{efron1994introduction} to estimate the standard deviation of our AUC estimates, by resampling the genetic risk scores or genetic risk predictions. For GeRSI, this is a straightforward bootstrap scheme. Denote $AUC(r_1,...,r_n,y_1,...,y_n)$ the estimated AUC given risk predictions $r_1,...,r_n$ and phenotypes $y_1,...,y_n$.
We sample with repetition $n$ indices $i_1,...,i_n \in \{1,...,n\}$. The $j$th bootstrap AUC sample is then:
\[AUC_j = AUC(r_{i_1},...,r_{i_n},y_{i_1},...,y_{i_n}),\]
and we report the empirical standard deviation of $100$ such bootstrap samples. 

When dealing with fixed-effects models, we must account for the fact that the p-value threshold is selected based on the AUC. The risk score of the $i$th individual, obtained when using $c$ as a threshold is denoted $r_{i}^{c}$. A bootstrap sample accounting for threshold selection is thus given by:
\[AUC_j = \underset{c}{\max}\{AUC(r^{c}_{i_1},...,r^{c}_{i_n},y_{i_1},...,y_{i_n})\}.\]

Lastly, when testings whether one method performs better than another, we note that comparing AUCs using estimated standard deviations is considerably conservative, since AUCs obtained using the same set of bootstrap indices are expected to be highly correlated. Instead, we follow a similar scheme to estimate the standard deviation of the difference in AUC directly.  

\section*{Estimating relative risk}
We are interested at estimating the relative risk of individuals at the top X of the risk predictions. A subtle aspect is that we wish to do so using case-control data, since we don't have a random sample from the population.
For a given risk threshold $v$, we estimate the fraction of the population with risk predictions higher than this threshold as $p_{pop}=\frac{K}{n_{cases}}\sum_{y_i=1}\mathbb{I}\{r_{i}>v\}+\frac{1-K}{n_{controls}}\sum_{y_i=0}\mathbb{I}\{r_{i}>v\}$. We then search for a value $v$ such that $p_{pop}$ is the desired value. 
The fraction of cases with risk predictions higher than the threshold is $p_{cases}=\frac{1}{n_{cases}}\sum_{y_i = 1 }\mathbb{I}\{r_{i}>v\}$, and the relative risk is estimated as: $\frac{p_{cases}}{1-p_{cases}} \times \frac{1-p_{pop}}{p_{pop}}$.

\section*{Population parameters }

Table \ref{pop_params_table} summarizes the values of prevalence and heritability which
were used for the estimation. 

\begin{table}[H]
\centering
\begin{tabular}{|c|c|c|}
\hline 
Phenotype & $K$ (\%) & $h^{2}$ (\%)\tabularnewline
\hline
\hline 
BD & 0.5 & 60\tabularnewline
\hline 
CD & 0.1 & 70\tabularnewline
\hline 
T1D & 0.5 & 80\tabularnewline
\hline 
T2D & 3 & 70\tabularnewline
\hline 
CAD & 3.5 & 50\tabularnewline
\hline 
RA & 0.75 & 60\tabularnewline
\hline 
HT & 5 & 60\tabularnewline
\hline
\end{tabular}
\caption{Population parameters used for estimating the performance of risk-prediction
methods. }
\label{pop_params_table}
\end{table}

\section*{Matrix identities for fast computation of the conditional mean and
variance }

This section contains some matrix identities which are used in the
software for faster computation. The first result is from \cite{stackExchange}

Let $\bar{x}\sim MVN(\bar{\mu},\Sigma).$ We are interested in the
conditional distribution $x_{i}\mid x_{-i}$. It is well known that
for the MVN case it is given by:

\[
x_{i}\mid x_{-i}\sim N(\Sigma_{i,-i}(\Sigma_{-i,-i})^{-1}(x_{-i}-\mu_{-i}),\Sigma_{ii}-\Sigma_{i,-i}(\Sigma_{-i,-i})^{-1}\Sigma_{-i,i}).\]

We wish to compute the mean and variance fast for every $i$. Naively,
this implies inverting an $(n-1)\times(n-1)$ matrix, for $n$ individuals
in the train set, resulting in running time complexity of $o(n^{4})$,
which is infeasible for realistic values of $n$. We are therefore
interested in computing the conditional mean and variance for every
individual in a more effective fashion. 

To do so we write (focusing on $i=n$ WLOG):

\[
\Sigma=\Big(\begin{array}{cc}
A & b\\
c^{\intercal} & d\end{array}\Big),\]

and \[
\Sigma^{-1}=\Big(\begin{array}{cc}
E & f\\
g^{\intercal} & h\end{array}\Big),\]

where $A$ and $E$ are $(n-1)\times(n-1)$ matrices so $b,c,f,g$
are column vectors and $d,h$ are scalars. 

The first (well known) result is that:

\[
A^{-1}=E-\frac{fg^{\intercal}}{h},\]

in other words, the inverses of all principal submatrices can be computed
from the inverse of the overall matrix. The spirit of the proof is
to note that:\[
\Big(\begin{array}{cc}
A & b\\
c^{\intercal} & d\end{array}\Big)\Big(\begin{array}{cc}
E & f\\
g^{\intercal} & h\end{array}\Big)=\Big(\begin{array}{cc}
I & 0\\
0 & 1\end{array}\Big),\]

hence $AE+bg^{\intercal}=I,$ and $A^{-1}=E(I-bg^{\intercal})^{-1}$.
By the Sherman-Woodbury-Morrison formula, we get \[
(I-bg^{\intercal})^{-1}=I+\frac{bg^{\intercal}}{1-g^{\intercal}b},\]

so \[
A^{-1}=E+\frac{(Eb)g^{\intercal}}{1-g^{\intercal}b},\]

Moreover, we can also write:

\[
\Big(\begin{array}{cc}
E & f\\
g^{\intercal} & h\end{array}\Big)\Big(\begin{array}{cc}
A & b\\
c^{\intercal} & d\end{array}\Big)=\Big(\begin{array}{cc}
I & 0\\
0 & 1\end{array}\Big),\]
from which we get:

\[
Eb+fd=0,\]

and:

\[
g^{\intercal}b+hd=1,\]

so plugging these into the previous result we get:

\[
A^{-1}=E-\frac{fg^{\intercal}}{h}.\]

However, we are not interested directly in $A^{-1},$ but in the conditional
mean and variance.

The conditional mean is given by $\Sigma_{i,-i}(\Sigma_{-i,-i})^{-1}(x_{-i}-\mu_{-i})$,
so all we need to compute is $\Sigma_{i,-i}(\Sigma_{-i,-i})^{-1}$,
in other words: \[
c^{\intercal}A^{-1}=c^{\intercal}(E-\frac{fg^{\intercal}}{h})=c^{\intercal}E-\frac{c^{\intercal}fg^{\intercal}}{h}=-dg^{\intercal}-\frac{(1-dh)g^{\intercal}}{h}=-\frac{g^{\intercal}}{h},\]

where we used the identities:

\[
c^{\intercal}E+dg^{\intercal}=0,\]

and

\[
c^{\intercal}f+dh=1.\]

Using our notation, the variance is:

\[
\Sigma_{ii}-\Sigma_{i,-i}(\Sigma_{-i,-i})^{-1}\Sigma_{-i,i}=d-c^{\intercal}A^{-1}b=d-c^{\intercal}(E-\frac{fg^{\intercal}}{h})b\]

Using again the fact that $c^{\intercal}f+dh=1$, and $c^{\intercal}E+dg^{\intercal}=0$
we get:

\[
...=d-c^{\intercal}(E-\frac{fg^{\intercal}}{h})b=d-(c^{\intercal}E-\frac{c^{\intercal}fg^{\intercal}}{h})b=d-(-dg^{\intercal}-\frac{(1-dh)g^{\intercal}}{h})b\]

\[
=d+\frac{g^{\intercal}}{h}b=d+\frac{1-dh}{h}=\frac{1}{h}.\]

\bibliographystyle{unsrt}
\bibliography{references}